\documentclass{aa}

\usepackage{gensymb} 
\usepackage{hyperref}

\usepackage{color}

\newcommand{\ha}{H_\mathrm{A}}
\newcommand{\hb}{H_\mathrm{B}}

\begin{document} 

	 \title{\textbf{ODEA}: Orbital Dynamics in a complex Evolving Architecture}
	\subtitle{Application to the planetary system HD 106906}

   	\author{L. Rodet\inst{1}, H. Beust\inst{1}, M. Bonnefoy\inst{1}, R. J. De Rosa\inst{2}, P. Kalas\inst{3}, A.-M. Lagrange\inst{1}}
   	
   	\institute{Univ. Grenoble Alpes, CNRS, IPAG, F-38000 Grenoble, France\\ 
    	     	\email{laetitia.rodet@univ-grenoble-alpes.fr} 
    	     	\and Kavli Institute for Particle Astrophysics and Cosmology, Stanford University, Stanford, CA 94305
    	     	\and Department of Astronomy, University of California, Berkeley, CA 94720 }

   	\date{Received --; accepted --}

  	\abstract{Mixed-variable symplectic integrators are widely used in orbital dynamics. However, they have been developed for Solar system-type architectures, and can not handle evolving hierarchy, in particular in systems with two or more stellar components. Such configuration may have occurred in the history of HD 106906, a tight pair of F-type stars surrounded by a debris disk and a planetary-mass companion on a wide orbit.}
  	{We present the new algorithm \texttt{ODEA}, based on the symplectic algorithm \texttt{Swift HJS}, that can model any system (binary,...) with unstable architecture. We study the peculiar system HD 106906 as a testcase for the code.}
  	{We define and compute a criterion based on acceleration ratios to indicate when the initial hierarchy is not relevant anymore. A new hierarchy is then computed. The code is applied to study the two fly-bys that occurred on system HD 106906, recently evidenced by De Rosa \& Kalas (2019), to determine if they could account for the wide orbit of the planet. Thousands of simulations have been performed to account for the uncertainty on the perturbers coordinates and velocities.}
  	{The algorithm is able to handle any change of hierarchy, temporary or not. We used it to fully model HD 106906 encounters. The simulations confirm that the fly-bys could have stabilized the planet orbit, and show that it can account for the planet probable misalignment with respect to the disk plane as well as the disk morphology. However, that requires a small distance at closest approach  ($\lesssim$ 0.05 pc), and this configuration is not guaranteed.}
  	{\texttt{ODEA} is a very good choice for the study of non-Solar type architecture. It can now adapt to an evolving hierarchy, and is thus suitable to study capture of planets and dust. Further observations of the perturbers, in particular their radial velocity, are required to conclude on the effects of the fly-by on system HD 106906.} 
 
 	
 	\titlerunning{Orbital dynamics in a complex evolving architecture}
 	\authorrunning{L. Rodet et al.}

 	\maketitle
 	 
	\section{Introduction}

	\subsection{Symplectic algorithms}
	
	In the context of the rapid increase of exoplanet discoveries, the need for efficient N-body simulations has become strong to model the evolution of complex systems and the interaction between planets, planets and debris disk, or within debris disks. Mixed variable symplectic integrators are widely used for dynamical simulations of planetary systems, as they present two major advantages with respect to other N-body integrators: First, they exhibit no long-term accumulation of energy error, which is essential to ensure orbital stability through the integration. On the other hand, they provide a gain of at least one order of magnitude in computation speed, for equivalent accuracy, because they allow one to adopt a much larger time-step than other integrators for the same result. In 1991, Wisdom and Holman devise the first symplectic map specifically designed for N-body problems with a central dominant mass \citep{wisdom1991}. Since then, numerous codes implemented this structure that are still widely used today (e.g., \texttt{Swift}, \citeauthor{levison1994} 1994, Mercury \citeauthor{chambers1999} 1999).
	
	Yet, symplectic integrators can model the interactions between multiple stars, moon, or simply planets whose mass are non negligible with respect to the central mass as well. They are versatile tools well suited to characterize the great diversity of extrasolar system architectures, well beyond the framework of our Solar System. Efforts were made to extend the scheme to binary stars in two modified versions of Mercury \citep{chambers2002}, but it could not be generalized to multiple systems with other hierarchies. In this context, \cite{beust2003} designed a symplectic scheme valid for any type of hierarchical architecture, and implemented it with \texttt{Swift HJS}. This generalized the theoretical frame of Wisdom and Holman to any hierarchical system. 
	
	However, in \texttt{Swift HJS}, the hierarchical structure of the system is given at the beginning of the run and must be preserved along the integration. This is a severe limitation as it prevents the efficient modeling of non stable hierarchies with e.g. orbital captures (planets, dust), whereas such situations may be numerous among young systems. With \texttt{Swift HJS}, handling accurately such configurations is only possible adopting a very small time-step, which is of course not optimal. This motivated us to build a new version of \texttt{Swift HJS}, \texttt{ODEA}, that tackles this issue. The code is available on \url{https://github.com/LaRodet/ODEA}.
	
In the following, we describe the new code in detail, and present a full application to the complex system of HD 106906. Before that, we present this system and our motivations for modeling it and using it as a benchmark for our new code.

	\subsection{HD 106906}
	
	The system HD 106906 (HIP 59960) is located at a distance of $103.3 \pm 0.5$ pc \citep{GaiaDR2} and belongs to the Lower Centaurux Crux (LCC) group, which is a subgroup of the Scorpius-Centaurus (Sco-Cen) OB association \citep{dezeeuw1999}. The LCC group has a mean age of $15 \pm 3$ Myr, with an age spread of 6 Myr \citep{pecaut2016}. HD 106906 is a $2.58 \pm 0.04~\mathrm{M_\odot}$ spectroscopic binary star, on an eccentric ($0.66$) and tight ($0.6$ au) orbit \citep{lagrange2019}.  Moreover, high contrast imaging has revealed an asymmetric debris disk \citep{kalas2015,lagrange2016} and a giant planet on a wide orbit  \citep[projected separation from the binary:  $735 \pm 5$ au,][]{bailey2013}. At such a separation, the planet relative motion can not be detected with present imaging instruments on a reasonable time basis. The orbital inclination with respect to the plane of the disk is probably significant (20\degree), but a coplanar configuration cannot be excluded. The planet mass has been estimated at $11\pm 9$ M$_J$ mass from hot-start models by \cite{daemgen2017}.
	
	Two major scenarios compete for the formation of giant planets \citep[e.g.,][]{baruteau2016}. In the core accretion scenario, planets begin their formation with the growth of dust grains and the formation of planetesimals, that will slowly accrete each other to form terrestrial planets or planetary cores. On the other hand, the gravitational instability scenario is a faster process that is able to form giant planets at large separation from an instability in the protoplanetary disk. In both cases, planet formation takes place in the primordial gaseous disk. Forming a giant planet at 700 au or more from any central star appears very unlikely in any of those scenarios, first due to the lack of circumstellar gas at that distance, and second because the corresponding formation timescale would exceed the lifetime of the gaseous disk. This led \cite{rodet2017} to propose a dynamical scenario to account for the planet's current separation. The scenario involves a traditional planetary formation within the gaseous disk, an inward migration and a subsequent scattering by the binary. However, for the planet to remain bound, an external perturbation such as a fly-by is necessary in order to reduce its eccentricity and stabilize its orbit in a bound configuration.
	
	Recently, \cite{derosa2019} investigated the stellar neighborhood of system HD 106906 in Gaia DR2 \citep{GaiaDR2}, and discovered two stars that have recently come within 1\,pc of the central binary HD 106906 AB. Given the uncertainty on the perturbers distances and radial velocities, De Rosa \& Kalas concluded that there was a possibility that the fly-by was dynamically significant for the planet evolution history. This motivates us to reinvestigate the \cite{rodet2017} scenario, using \texttt{ODEA}, to check this possibility.
	
	\section{Algorithm}
	
	\subsection{Structure of the code: \texttt{Swift HJS}}
	\label{sec:structure}
	
	Let us consider the gravitational N-body problem, with masses $(m_i)_{i=1,..,N}$,  positions $(\vec{r}_i)_{i=1,..,N}$ and impulsions $(\vec{p}_i)_{i=1,..,N}$. The Hamiltonian is
	
	 \begin{equation}
 	 	H = \sum_{i=1}^N {\frac{\vec{p_i}^{2}}{2m_i}} - \sum_{1\leq i<j \leq N} {\frac{G m_i m_j}{r_{ij}}} \quad ,
 	 \end{equation}
 	 
 	 \noindent where $G$ is the constant of gravitation and $r_{ij} = ||\vec{r}_j - \vec{r}_i||$ is the distance between bodies $i$ and $j$.
	
	In the current version of \texttt{Swift HJS}, as in the other similar codes, the integrator do not solve $H$ exactly, but a surrogate Hamiltonian $\tilde{H}$. The latter is chosen to be close to the real one, and exactly solvable. In that case, the algorithm is symplectic: it exactly preserves the areas in phase space and exhibit no long-term drift of the energy.
	
	In order to design a proper $\tilde{H}$ in orbital mechanics, the key idea is to split the Hamiltonian into two integrable parts:
	 	
 	 \begin{equation}
 	 	H = \ha + \hb \quad .
 	 \end{equation}
 	 
 	 Several splitting have been suggested \citep[e.g., ][]{wisdom1991,saha1994,chambers1999}, most of them consisting on a Keplerian part and a perturbation part. Both parts are then integrable within computer round-off errors. $\tilde{H}$ corresponds to the successive integration of these parts separately. For a second order symplectic integrator, a so-called leap-frog method can be used. It consists in integrating $\hb$ for $\Delta t/2$ (kick), then $\ha$ for $\Delta t$ (drift), then again $\hb$ for $\Delta t/2$ (kick), where $\Delta t$ is the time step.
 	 
 	  \texttt{Swift HJS} is based on the Hierarchical Jacobi Symplectic method introduced by \cite{beust2003}, where the description is based on orbits instead of on bodies. An orbit consists in a collection of two non-empty sets of bodies, the set of centers and the set of satellites, that have empty intersection. In all problems in orbital mechanics, a hierarchy can then be defined as a collection of orbits comprising all bodies satisfying the following rule: for all couples of orbit $k$ and $l \neq k$, one of the three subsequent propositions apply

	\begin{itemize}
		\item orbits k and l have no common bodies  (orbits $k$ and $l$ are \textit{foreign});
		\item orbit $k$ is comprised in the centers or satellites of orbit $l$ (orbit $k$ is \textit{inner} to orbit l);
		\item orbit $l$ is comprised in the centers or satellites of orbit $k$  (orbit $k$ is \textit{outer} to orbit l).
	\end{itemize}
	
	A so-defined hierarchy is made of exactly $N-1$ orbits. In \texttt{Swift HJS}, the orbits are numbered from 2 to N. Finally, we define $\mu_k$ and $\eta_k$ as the total mass of the satellites and centers respectively in orbit $k$. The total dynamical mass in orbit $k$ is then $M_k = \mu_k + \eta_k$ and the reduced mass $m'_k = \mu_k \eta_k / M_k$. 
 	 
 	  In this formalism, a new set of $N$ coordinates $(\vec{r'}_k,\vec{p'}_k)_{i=1,..,N}$ are designed with a Jacobi-like approach: $\vec{r'}_k$ is the relative position of the center of mass of orbit $k$'s satellites with respect to that of its centers, and $\vec{p'}_k$ is the relative conjugate momentum. The first coordinates $\vec{r'}_1$ and $\vec{p'}_1$ are the position and impulsion of the center of mass. These positions and conjugate momenta derive from a canonical transformation that let the Hamiltonian invariant. They can be expressed with the bodies coordinates as
 	  
	\begin{align}
		\vec{r'}_k &= \sum_{i\text{, satellites of k}} \frac{m_i \vec{r}_i}{\mu_k} - \sum_{i\text{, centers of k}} \frac{m_i \vec{r}_i}{\eta_k}\\
		\vec{p'}_k &= m'_k \left(\sum_{i\text{, satellites of k}} \frac{\vec{p}_i}{\mu_k} - \sum_{i\text{, centers of k}} \frac{\vec{p}_i}{\eta_k}\right)
	\end{align}	 	  
 	  
 	 The Hamiltonian can then be split as follows
 	 
 	  \begin{align}
 	 	\ha &= \sum_{k=2}^N {\frac{{p'_k}^{2}}{2m'_k} - \frac{G\mu_k\eta_k}{r'_k}} \label{eq:ha} \quad ;\\
 	 	\hb &= \sum_{k=2}^N { \frac{G\mu_k\eta_k}{r'_k}} - \sum_{1\leq i < j < \leq N} { \frac{Gm_im_j}{r_{ij}}} \label{eq:hb} \quad .
 	 \end{align}

 	 When the hierarchy is sufficiently clear (that is if the orbits are almost Keplerian), $\hb \ll \ha$. As $\ha$ is a Keplerian Hamiltonian describing $N-1$ independent orbits, the drift consists of evolving each Keplerian orbits. On the other hand, as $\hb$ depends exclusively on the positions, the kick consists of a linear raise of the velocities, with accelerations $\vec{a_k^B} \equiv 1/m'_k~\partial \hb / \partial \vec{r'_k}$.
 	 
 	 \subsection{Building a new hierarchy}
		
\begin{figure}
		\includegraphics[width=\linewidth]{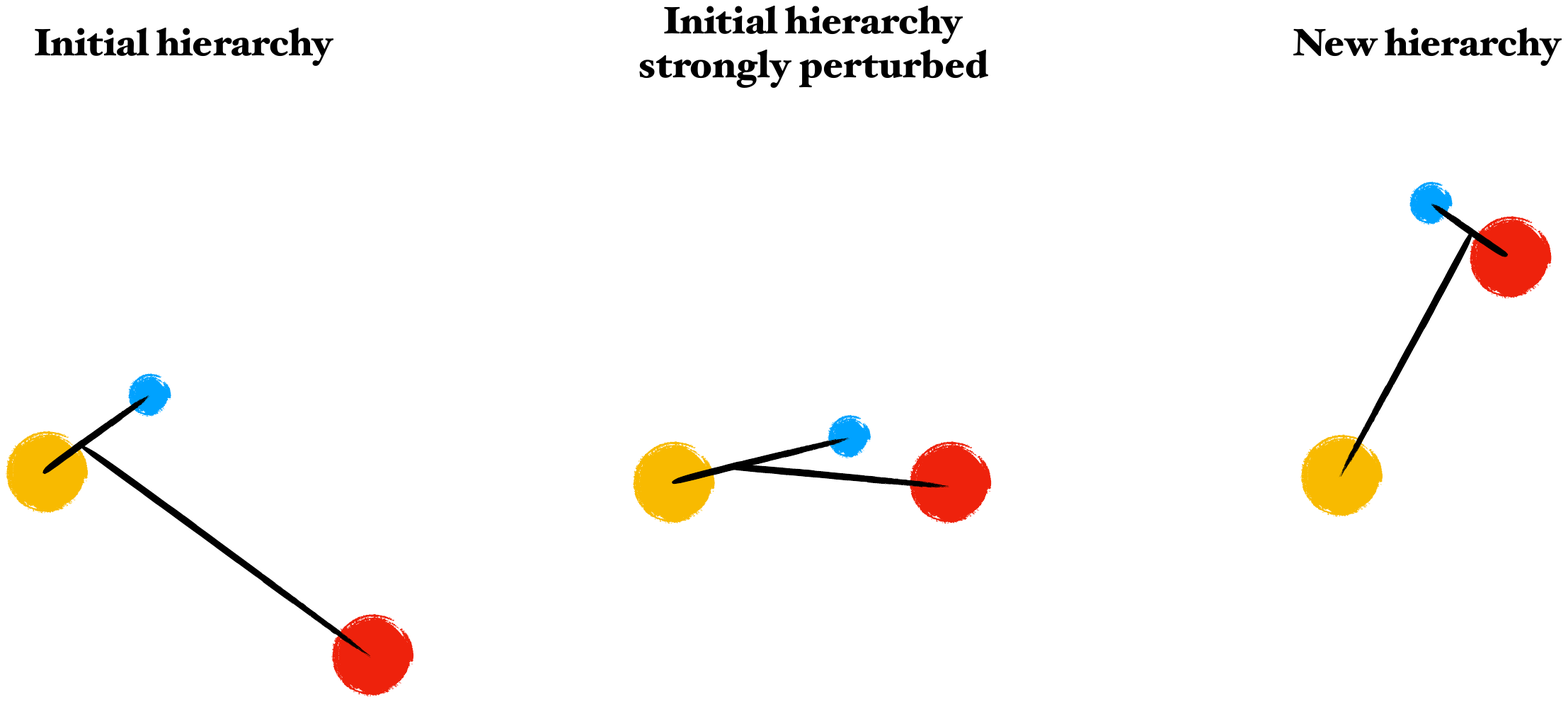}
		\caption{Example of hierarchy change in the case of a capture. At first the red body orbits the yellow--blue pair. After a strong interaction, it captures the small blue body.}\label{fig:capture}
	\end{figure}

	The above scheme is well adapted to lightly perturbed Keplerian orbits in a fixed hierarchy, but becomes strongly unsuitable if the initial hierarchy evolves, whether temporarily or definitively (see example Fig. \ref{fig:capture}).
	
	Thus, when the hierarchy is not relevant anymore (that is the splitting in the initial $\ha$ and $\hb$ does not optimize the error), a module of the algorithm will design a new hierarchy from the current positions of the bodies. For this, the algorithm computes a two-dimensional symmetric array that compiles the Keplerian acceleration between two bodies $a_k^\text{Kep} = GM_k/r_{ij}^2$, where $M_k$ is the sum of the masses. The strongest acceleration gives the first orbit, then the two bodies are replaced by their center of mass and the array is updated, and again until the last orbit comprises all bodies. We first checked that this algorithm always returns the existing hierarchy when no change is expected. Then, if the computed hierarchy is different than the current one, the hierarchy must be changed.
	
	
	If the hierarchy needs to be changed, so is the time-step $\Delta t$. We choose a Keplerian-like time $\min_k T_k/20$, where
	
	\begin{equation}
		T_k = \sqrt{\frac{4\pi^2a_k^3 \mid 1-e_k\mid ^3}{GM_k}}
	\end{equation}
	
	\noindent if orbit $k$ is bound or if its smallest approach has not yet occurred, or
	
	\begin{equation}
		T_k = \sqrt{\frac{4\pi^2 {r'_k}^3}{GM_k}}
	\end{equation}
 	 
 	 \noindent otherwise. The choice to adapt or not the time step is given to the user.
	
	Strictly speaking, when changing the hierarchy, the symplectic nature of the algorithm does not hold anymore, as the splitting of the Hamiltonian is entirely based on the hierarchy. This is also true for any change of the time step. A new approximate Hamitonian is integrated from an already approximated scheme, which means that the error budget raises potentially at each hierarchy change. However, the algorithm is designed for orbital dynamics, where systems are not subject to frequent reorganization of their architecture. Designing a new Hamiltionan when the initial hierarchy is not suited anymore allows to limitate the error on each orbit, which will otherwise become out of control. This is basically the same problem as the one raised by close encouters in planetary dynamics. When handling close encounters, \cite{levison1994} (in \texttt{Swift RMVS}) and \cite{chambers1999} (in \texttt{Mercury}) temporarily change the way of splitting the Hamiltonian when transferring to $\ha$ the part of $\hb$ that concerns the close encounter, even sometimes changing the hierarchy to planetocentric (in the latest version of \texttt{Mercury} the use of a smooth criterion that weights the different perturbing terms allows the map to remain symplectic with a continuous Hamiltonian while handling close encounters; \citeauthor{rein2019} \citeyear{rein2019}).  Conceptually, a close encounter within a planetary system can be viewed as a temporary change of hierarchy that eventually returns to the initial hierarchy. Here we are concerned by changes that can be permanent.
	
	\subsection{Checking the relevance of the hierarchy}
	
	Performing a hierarchy change is quite costly: all the acceleration couples have to be computed and must be compared and updated for the definition of each of the $N-1$ orbits (multiple operations that scale as $O(N^3)$). Checking for a possible change at each time-step, with the result that most of the time the current hierarchy would be left unchanged, would thus amount to a considerable loss of efficiency. Prior to launching the entire hierarchy re-building process, an efficient algorithm with a simpler criterion must be applied to check whether it is appropriate or not. The most exact criterion would be the theoretical energy error associated to the symplectic mapping, as it gives us an objective estimate of the relevance of the numerical scheme. However, its computation is tedious (grows as $N^4$, see Appendix). The criterion must be fast to compute (maximum as $N^3$, like the accelerations) and correlated to the error.
	
In Mercury \citep{chambers1999}, the criterion to spot close encounters is the ratio between the relative distances and the Hill radii, assuming the latter roughly constant. This is a legit criterion for the study of the Solar system, but it is not relevant to our case. Indeed, the Hill radius is not easy to compute for eccentric orbit, it depends strongly on the orbital parameters (which is subject to variation in the general case) and it is not satisfyingly correlated to the errors in a complex architecture.
	
	We choose to compute at each step the ratio $a^B_k/a^\text{Kep}_k$ for each orbit $k$, where $a^\text{Kep}_k$ and $a^B_k$ are the accelerations $\ddot{\vec{r'_k}}$ respectively induced by $\ha$ and $\hb$ (Eqs \ref{eq:ha} and \ref{eq:hb}). We declare the hierarchy questionable if it is higher than $0.2$ for at least one orbit. The computation of that criterion also scales as $O(N^3)$ in theory, but it uses the acceleration $a^B$ that is already computed in any step of the integration, so that the extra cost remains limited. 
	
	Thus, with this criterion, the problem can keep a non-optimal hierarchy if the associated error remains small. This can be adjusted by changing the value of the threshold, which is a free parameter of the code. This might be useful in situations when two similar hierarchies become alternately optimal, to prevent the algorithm to perform numerous changes that will have a negative effect on long term conservation properties.
	
	\subsection{The case of test particles}
	
	The study of planetary systems often involved the study of debris belts. In N-body simulations, the dust is modeled at first order by massless bodies (or test particless) that interact with the massive bodies but not with each other. Test particles must be specifically considered in \texttt{ODEA} as the handling of their hierarchy is slightly different. Indeed, they are the only satellites of their orbit and their orbit is invisible to the bodies and other test particles evolution. When looking for a new hierarchy, \texttt{ODEA} will not consider the test particles, for it searches foremost to optimize the energy error budget related to the massive bodies.
	
	When the hierarchy of the massive bodies changes, each test particle must find its natural orbit given its relative position. A similar procedure to the hierarchy building of massive bodies is then performed. For a consistent hierarchy, the test particles have $2N-1$ possibilities for their orbit: around one massive bodies ($N$) or around one orbit ($N-1$). Thus, for each test particle, a $2N-1$ array is computed, compiling the Keplerian accelerations. The maximal element will correspond to the new particle configuration.
	
	Finally, a test particle may also be subject to a hierarchy change, independently of the massive bodies architecture evolution. Thus, the acceleration ratio criterion is computed at each time step to check the suitability of the particle orbit, and a new orbital configuration is investigated if necessary following the previous procedure.
	
	\subsection{Comparison with other codes}
	
	Several algorithms have been introduced since the formalization of the first mixed-variable symplectic map for orbital mechanics, including the widely used \texttt{Mercury} \citep{chambers1999}. Most of them are designed to work in Solar-System-like hierarchy. \cite{chambers2002} introduced two algorithms, derived from \texttt{Mercury}, to model planetary motions in binary systems. However, to our knowledge, no symplectic integrator are able to integrate indifferently any types of hierarchy, or a more complex hierarchy, except from \texttt{Swift HJS}.
	
	\begin{figure}
		\centering
		\includegraphics[width=\linewidth]{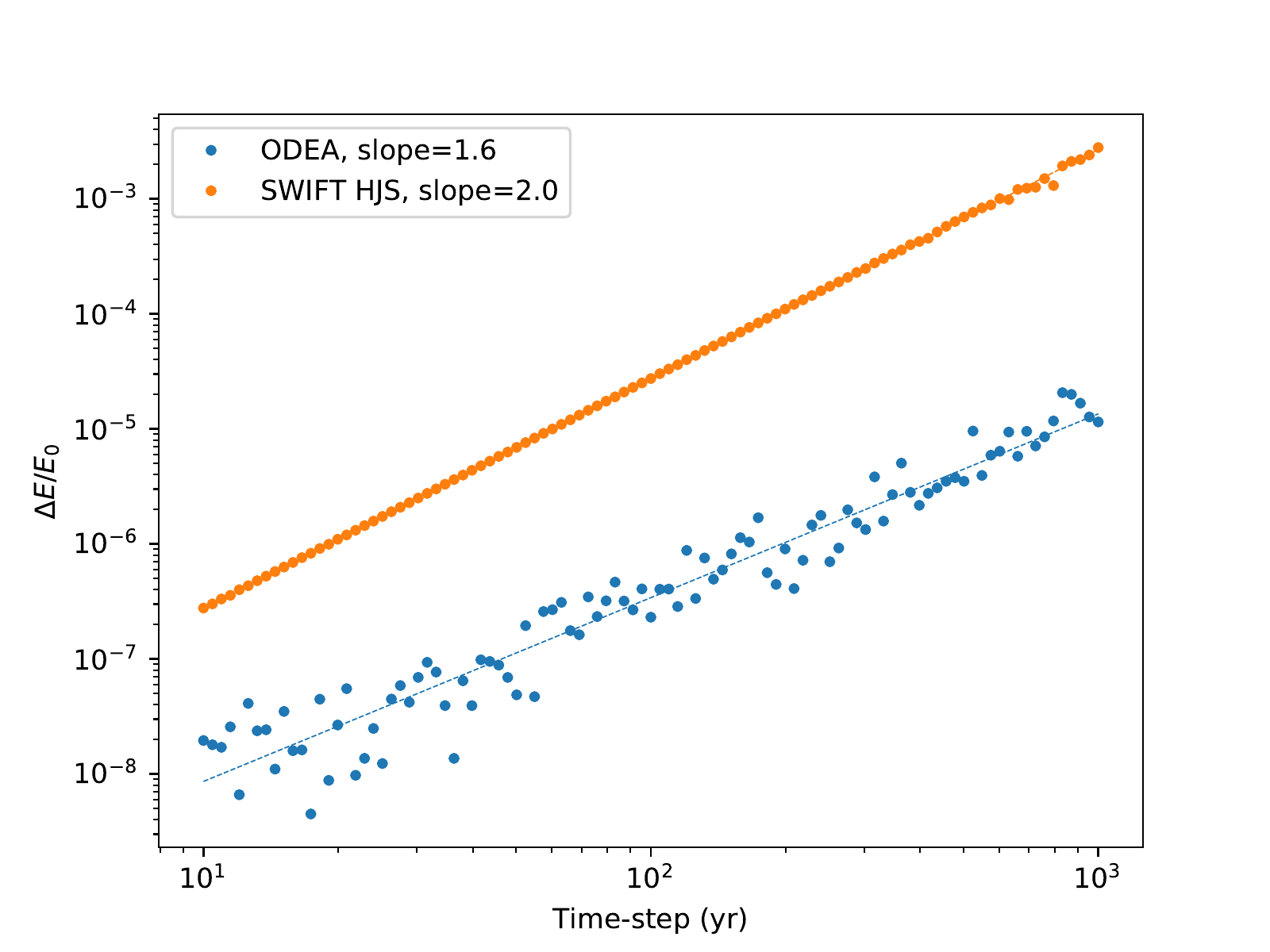}
		\caption{Maximum relative energy error over a 1 Myr evolution of HD 106906 including the two fly-bys, for \texttt{ODEA} and \texttt{Swift HJS}, as a function of the time-step assumed.}\label{fig:de}
	\end{figure}
	
	Moreover, no mixed-variable integrator that we know of are designed to handle long or definitive hierarchy change. Such situations can be encountered in case of a stellar fly-by, or of a capture of debris disk dust by a stellar or planetary companion. The subsequent study of system HD 106906 is a perfect example of situations that can not be tackled by ordinary symplectic algorithms: binary fly-by and dust capture. Fig. \ref{fig:de} illustrates the gain of energy precision that \texttt{ODEA} allows on the HD 106906's fly-bys test case (the parameters of the corresponding simulation are presented in the Appendix). The relative energy error is here entirely dominated by the close encounters. By changing the hierarchy, \texttt{ODEA} decreases the error of one or two orders of magnitude compared to \texttt{Swift HJS}. Moreover, it allows the energy error to decrease after the encounter in case of definitive hierarchy change.
	
	On an other hand, \cite{rein2014} argue that a high-order classical integrator is quicker and more accurate than symplectic integrators. This may be true for some complex cases, or if we aim for a very high precision. However, symplectic integrators have encoded the exact resolution of the Keplerian motion, while a classical integrator makes no hypothesis for the form of the motion, and has to solve from scratch the differential equations of motion. Thus, for lightly perturbed Keplerian motion, symplectic algorithms are certainly more practical than classical integrators. The time steps can be large without endangering the stability of the orbits.
	
	For example, in the case of HD 106906, the simulations involved very different scales, from the planet periastron to the wide hyperbolic orbit of the perturbers. A classical integrator would have to adapt its time step to the smallest distance, while a symplectic integrator can adopt a larger timescale without compromising the stability of the planet orbit. This is illustrated by Table \ref{fig:de}, where the two symplectic integrators \texttt{ODEA} and \texttt{Swift HJS} can achieve a reasonable precision with a large time step (same simulation than for Fig. \ref{fig:de}, but for the entire 15 Myr evolution). Outside the fly-bys, they reach a precision similar to that of the classical integrator \texttt{IAS 15}, that has to decrease regularly its time step to resolve the periastron passage, increasing the computation time. We also ran the simulation with the Bulirsch-Stoer implementation of \cite{press1989}, with a precision constraint on the trajectory of order $10^{-7}$ (similar to the value reached by \texttt{ODEA}). The energy error grows very rapidly, and the computation time is already significantly larger than the other codes.
	
We also point out that \texttt{Swift HJS} never makes the assumption that the orbits we are considering are actually bound. The only requirement is that the sum of the Keplerian interactions associated with the hierarchy (i.e. $H_A$) must represent most of the full Hamiltonian. Some of the \textit{orbits} we are considering can thus be hyperbolic, and this will be the case in a fly-by configuration. The Kepler solver used to integrate $H_A$ handles bound or unbound orbits as well.

	\begin{table}
		\caption{N-body simulations of the 15 Myr past evolution of HD 106906 including the two fly-bys. The time step has been fixed to 100 yr in \texttt{ODEA} and \texttt{Swift HJS}.}\label{table:de}
		\begin{tabular}{ccc}
		\hline
		Code & CPU time (s) & $\Delta E/E_0$\\
		\hline
		\texttt{Swift HJS} & 4 & $3.10^{-5}$ \\
		\texttt{ODEA} & 7 & $3.10^{-7}$ \\
		\texttt{IAS 15} (\texttt{Rebound}) & 320 & $3.10^{-9}$ \\
		Bulirsch-Stoer & 600 & $5.10^{-4}$\\
		\hline
		\end{tabular}
	\end{table}
	
	\section{Application to system HD 106906}
	
	\subsection{Characterizing the perturbers}
	
\begin{figure}
		\includegraphics[width=\linewidth]{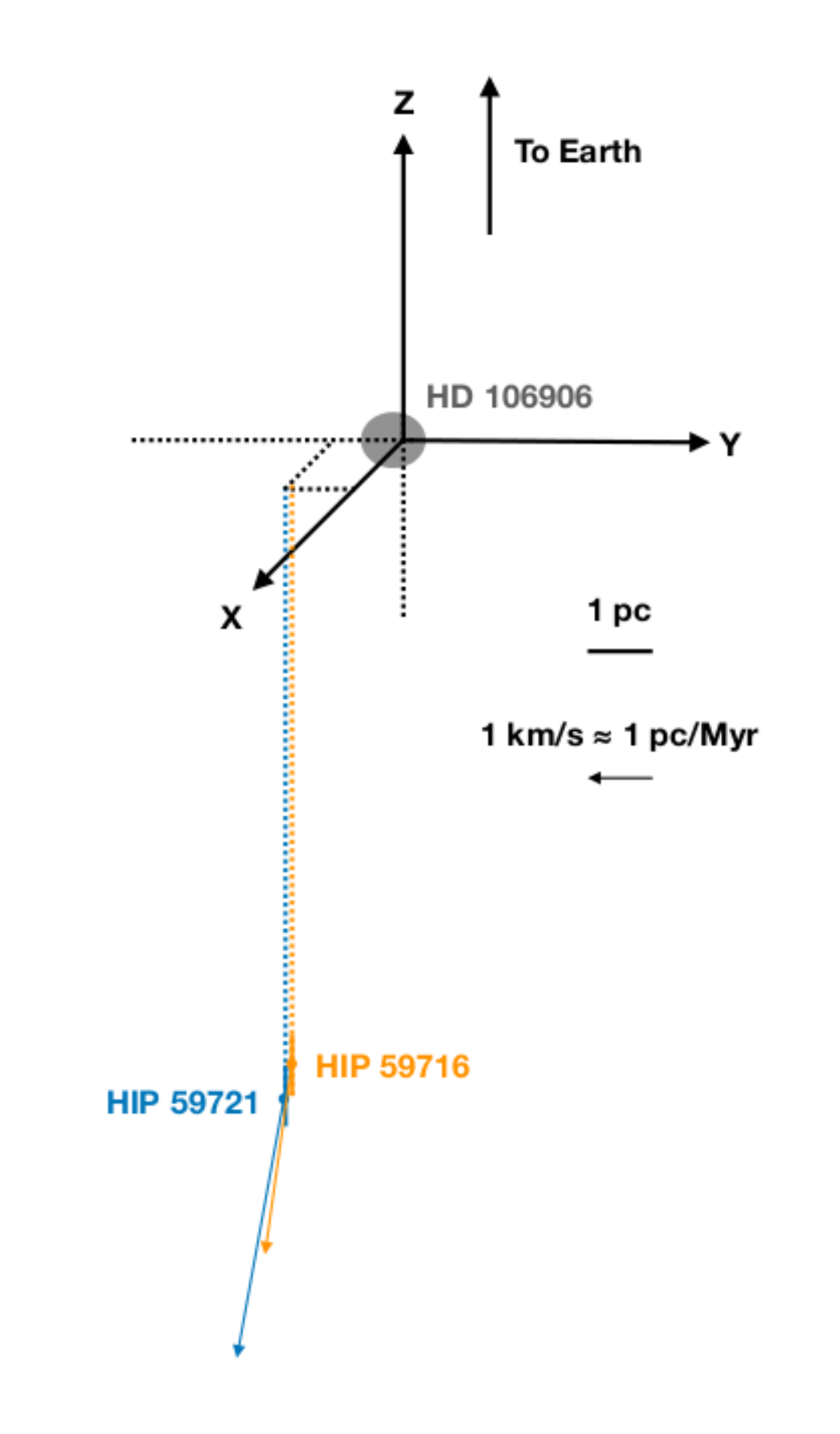}
		\caption{Representation of HD 106906, HIP 59716 and HIP 59721 current positions and velocities in HD 106906 rest frame (disk lies in the YZ plane, observed extension in the -Y direction). }\label{fig:schema}
	\end{figure}

\begin{figure}
		\includegraphics[width=\linewidth]{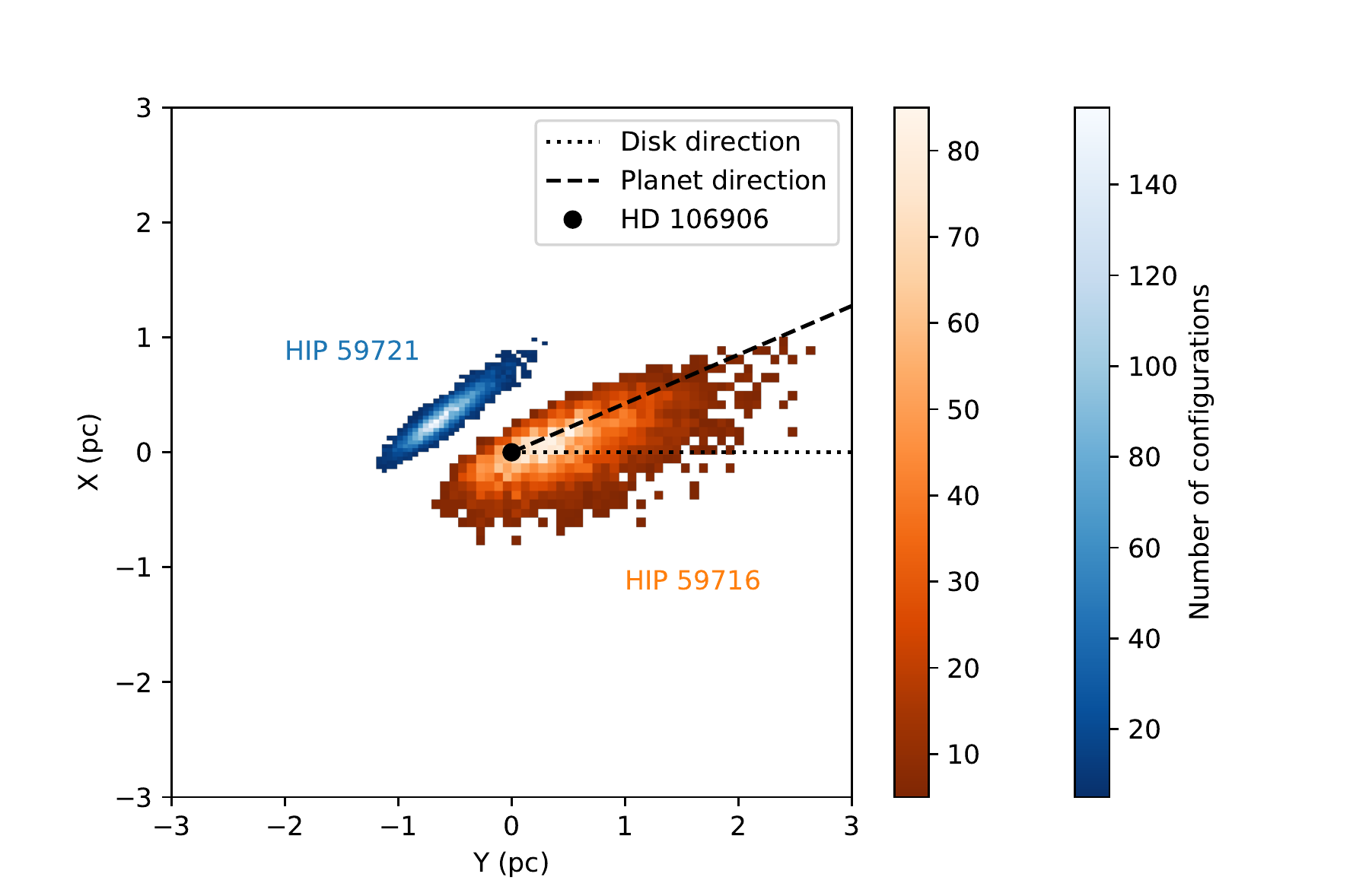}
		\caption{Two dimension histograms of the coordinates of the intersection points between the perturbers trajectories and the XY plane, assuming linear trajectories.}\label{fig:impact}
	\end{figure}

	Searching for potential stellar perturbers in Sco-Cen during the previous 15 Myr, \cite{derosa2019} identified two perturbers in LCC \citep{pecaut2012}: HIP 59716 and HIP 59721. Located around 11 pc (projected 0.5\degree) from HD 106906 and 0.5 pc (projected 30'') from each other, their relative velocities suggest an encounter with HD 106906 a few million years ago. The coordinates and velocities of the three systems are summarized in Table 1 of \cite{derosa2019}. As can be seen on Fig. \ref{fig:schema}, the relative separation and velocity between HD 106906 and its perturbers lie essentially on the direction to Earth. Unfortunately, the quantities projected in this direction (distance and radial velocity) have the larger observational uncertainties, which creates a high dispersion on the closest encounters, in particular for the most promising candidate HIP 59716 (Fig. \ref{fig:impact}).
	
	
	We note that the relative velocities between each systems ($\sim$ 4 km/s) are four times higher than the velocity dispersion reported for LCC \citep[$1.13 \pm 0.07$ km/s; ][]{madsen2002}, that was used in \cite{rodet2017}. We will see in subsection 3.3 that the effect of a fly-by is inversely proportional to the velocity of the passing star.
	
	The masses of HIP 59716 and HIP 59721 have been estimated respectively $1.37~\mathrm{M_\odot}$ for HIP 59716 and $1.22~\mathrm{M_\odot}$ for HIP 59721 from the spectral types. HD 106906 binary mass has been estimated to $2.58 \pm 0.04~\mathrm{M_\odot}$ from radial velocity and interferometric measurements by \cite{lagrange2019}.
	
	\subsection{Simulating the encounters}
	\label{sec:simulations}
	
	N-body simulations performed by \cite{derosa2019} indicate that the galactic gravitational potential has a negligible influence on the characteristics of the encounters. Moreover, the binarity of HD 106906 does not affect the encounters, because of the very high ratio between the closest approaches and the binary separation (> 1000). In order to efficiently determine the parameters of the encounters, we first performed $10,000$ simulations with \texttt{ODEA}, including three bodies: HD 106906 ABb ($2.58 + 0.01~\mathrm{M_\odot}$), HIP 59716 and HIP 59721. The mass of HD 106906 and the algorithm that we present here are the only differences with De Rosa \& Kalas study at that point.
	
	The initialization of the simulations is designed with a Monte-Carlo approach, following De Rosa \& Kalas. The $3\times 6$ parameters and their respective precision are the right ascension $\alpha$ (0.05 mas), the declination $\delta$ (0.002 mas), the parallax $\pi$ (0.05 mas), the proper motion of the right ascension $\mu_\alpha \cos\delta$ (0.05 mas/yr), the proper motion of the declination $\mu_\delta$ (0.05 mas/yr) and the radial velocity $\gamma$ (up to 1.7 km/s). The parameters are drawn from a normal distribution centered on their measured values, with a dispersion equal to the observations uncertainties, taking into account the correlations given by Gaia catalog. Then, we trace back the stars trajectory to observe the encounters.
	
\begin{figure}
		\includegraphics[width=\linewidth, trim = {0 6cm 0 0}]{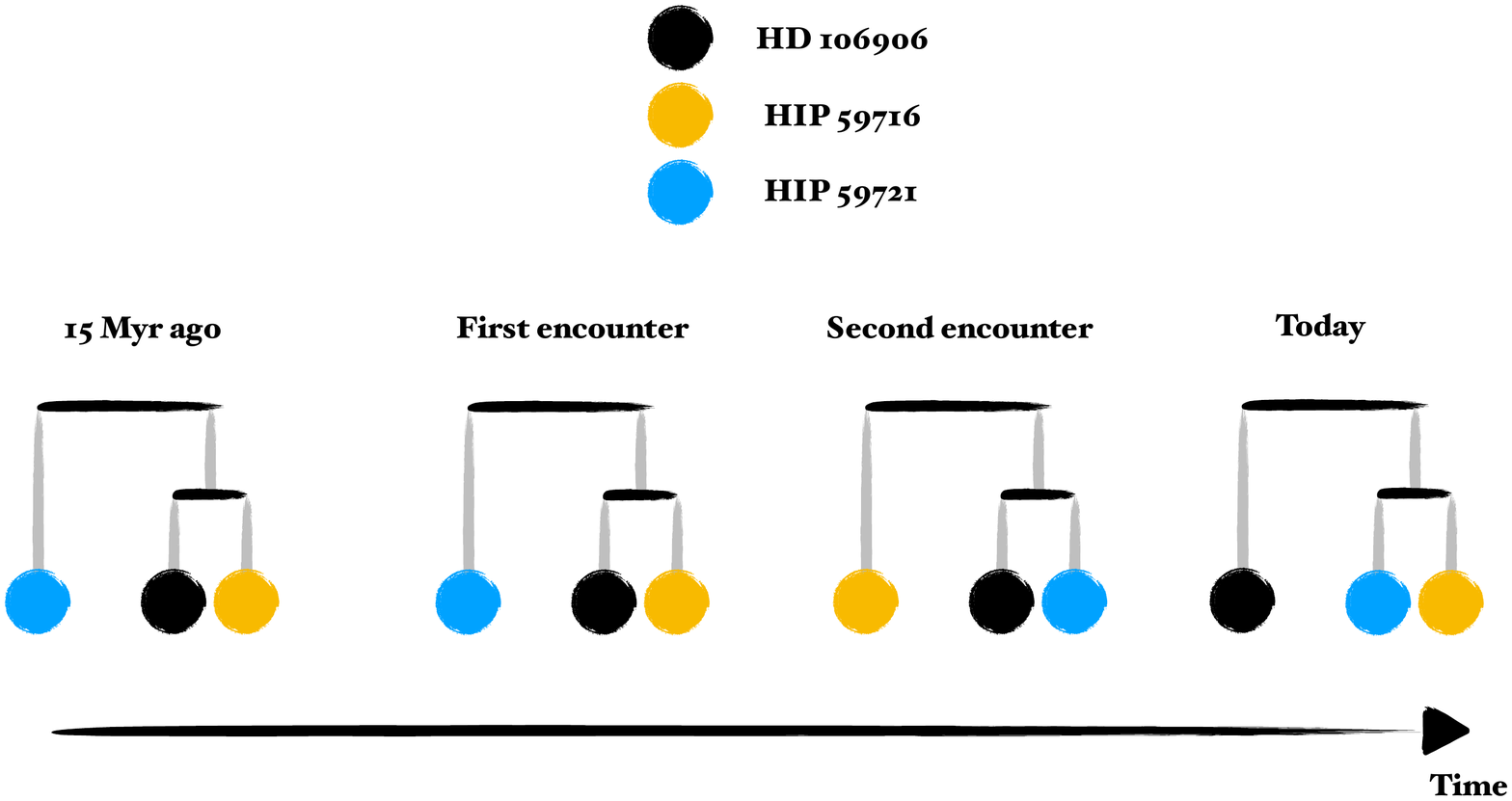}
		\caption{Representation of a typical evolution of the hierarchy in the three-body simulations of HD 106906 fly-bys with \texttt{ODEA}. All orbits here are hyperbolic.}\label{fig:hierarchy}
	\end{figure}

	Most of the simulations follow the same hierarchy evolution, represented on Fig. \ref{fig:hierarchy}: the first fly-by involves HIP 59716 and the second HIP 59721, before the two perturbers get very close at each other as can be seen today. The hierarchy will thus naturally evolves to take into account the successive encounters. Computing the eccentricity of several sets of configurations, we evaluated that the two perturbers have currently a $2.1 \pm 0.1$ \% chance of being gravitationally bound to each other. However, De Rosa \& Kalas point out that the probability of them having such similar angular positions and proper motions without being bound are extremely low.
	
	
	We launched 10,000 simulations for 15 Myr, corresponding to a backward evolution from our days to the formations of the stars. The different timescales of the simulations are summarized in Table. \ref{table:timescales}. At first sight, 10,000 simulations may not seem enough to correctly sample the 18 parameters confidence intervals. However, most of the parameters are strongly constrained, the only strong uncertainties being the perturbers relative radial velocities and distances, that is 4 parameters. Thus, these are the critical parameters that must be correctly sampled, and 10,000 is then a sufficient number. The initial time-step was set to 1,000 yr, with outputs every 1,000 yr. To account for the possibility of the two perturbers being bound, we performed an additional 10,000 simulations with only bound configurations. It comes down essentially to selecting only the configurations where the perturbers have similar radial velocities.
	
	\begin{table}
		\caption{Timescales of the HD 106906 simulations. }
		\begin{tabular}{cc}
		\hline
		Objects & Timescale (yr)\\
		\hline
		Host binary star period & $10^{-1}$ \\
		Planet period & $10^3$ \\
		Duration of the fly-by & $10^5$ \\
		Perturbers binary period & $10^{6}$ \\
		Time of the fly-by & $3.10^7$ yr ago\\
		Age of the system & $(15 \pm 6) .10^7$\\
		\hline
		\end{tabular}\label{table:timescales}
	\end{table}

	\begin{figure}
		\includegraphics[width=\linewidth]{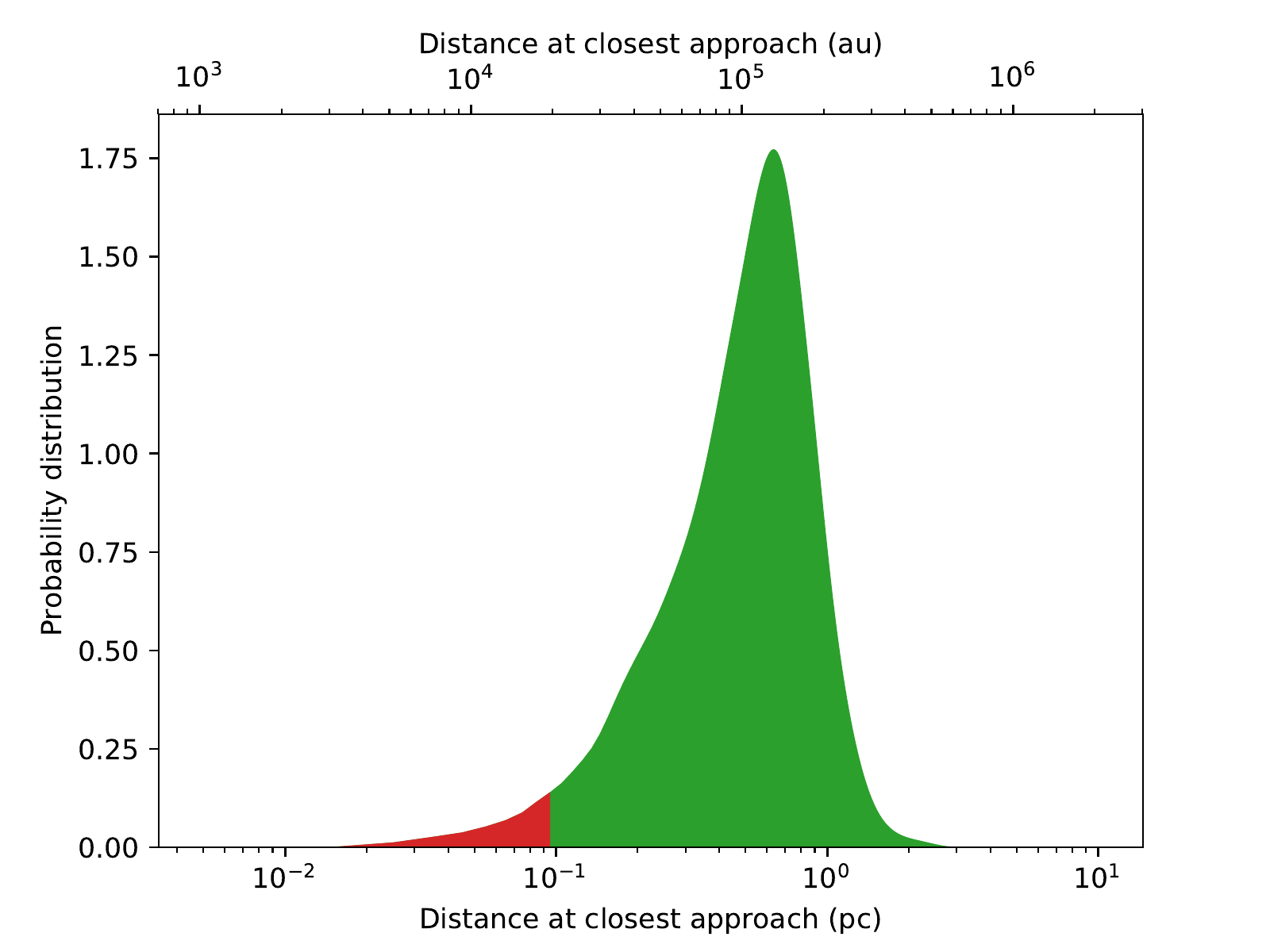}
		\caption{Distribution of the distances at closest approach. The following study will focus on the red part, that corresponds to fly-by closer than 0.1 pc (3.6 \% of the configurations).}\label{fig:dca}
	\end{figure} 
	
\begin{figure*}
		\includegraphics[width=\linewidth]{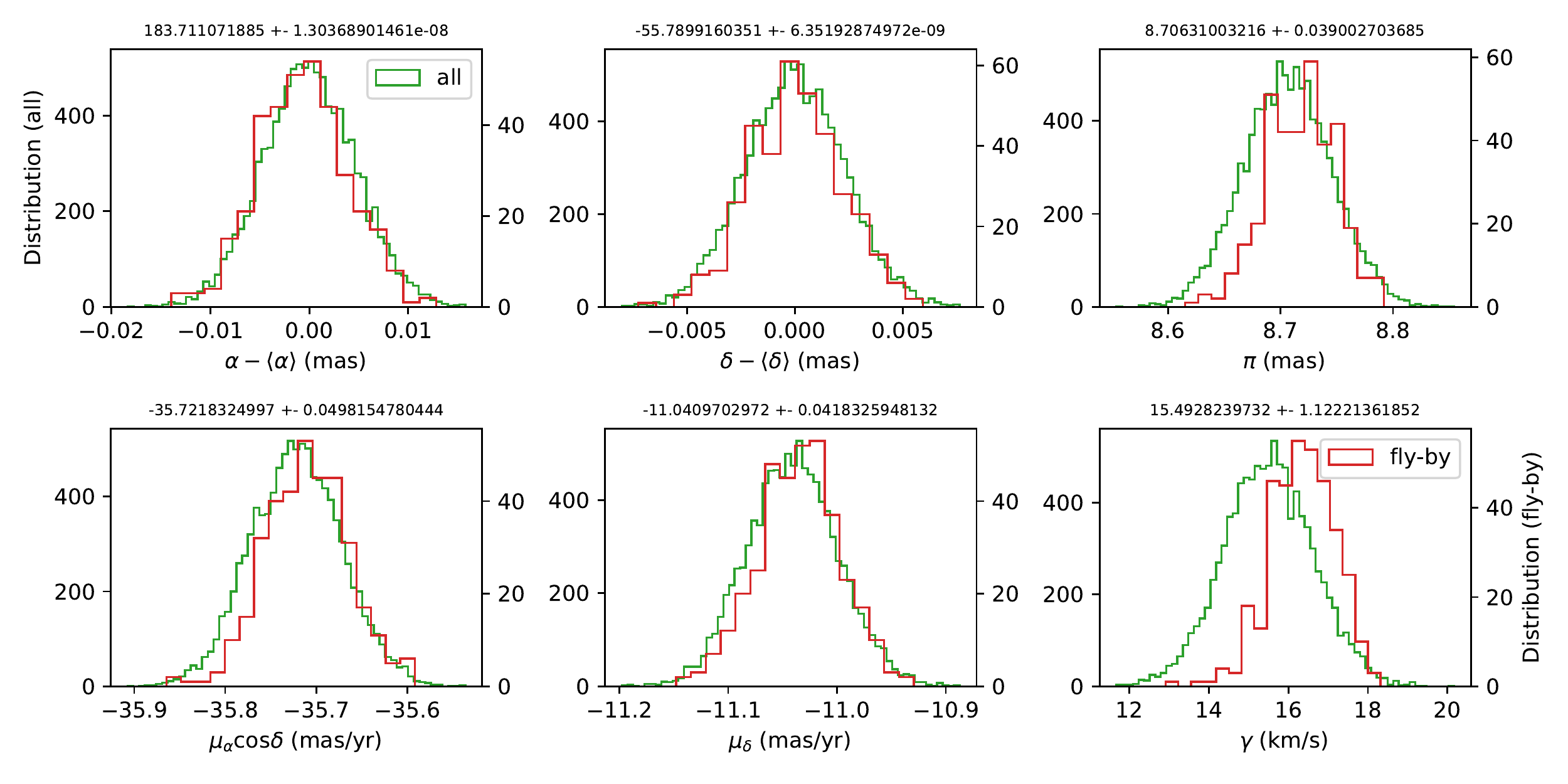}
		\caption{Initial distribution (today) of HIP 59716 coordinates and velocities for the 10,000 simulated cases (green), and for the 359 cases where a fly-by closer than 0.1 pc occurred (red).}\label{fig:coordinates2}
	\end{figure*}

	The distances at closest approach were computed for each simulation (Fig. \ref{fig:dca}). Most of the encounters occur with a closest approach between 0.3 and 2 pc, with a maximal probability around 0.6 pc, consistent with the results of De Rosa \& Kalas. We then reviewed the simulations for which a close (< 0.1 pc) fly-by occurred, from any one or both of the two perturbers. 359 configurations were selected, that is around 4\% of the total number of studied configurations. In most cases ($\gtrsim$ 90\%), HIP 59716 encounters HIP 106906 at the shortest distance. For the bound configurations, the peak is around 0.4 pc but the number of close fly-bys is roughly the same. HIP 59716 coordinates distributions are presented on Fig. \ref{fig:coordinates2}. Most of the parameters of the configurations with close fly-bys are drawn randomly within the configurations, except for the radial velocity, where we see that the configurations leading to a close fly-by correspond to the higher radial velocities (closer to the radial velocity of HIP 59721). The distributions for the two other bodies are presented on Fig. \ref{fig:coordinates1} and \ref{fig:coordinates3} in the appendix.
		
	\begin{figure}
		\includegraphics[width=\linewidth]{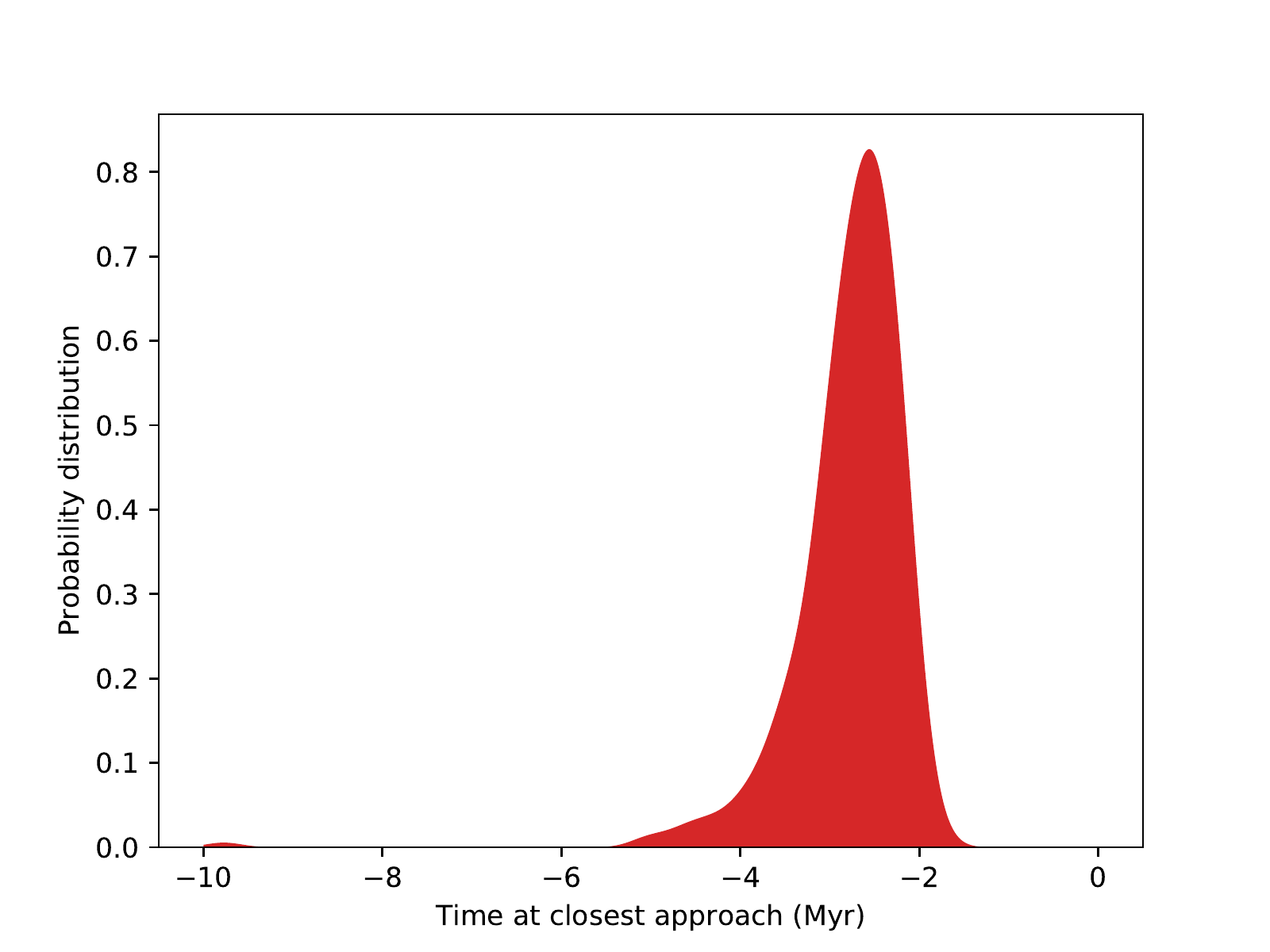}
		\includegraphics[width=\linewidth]{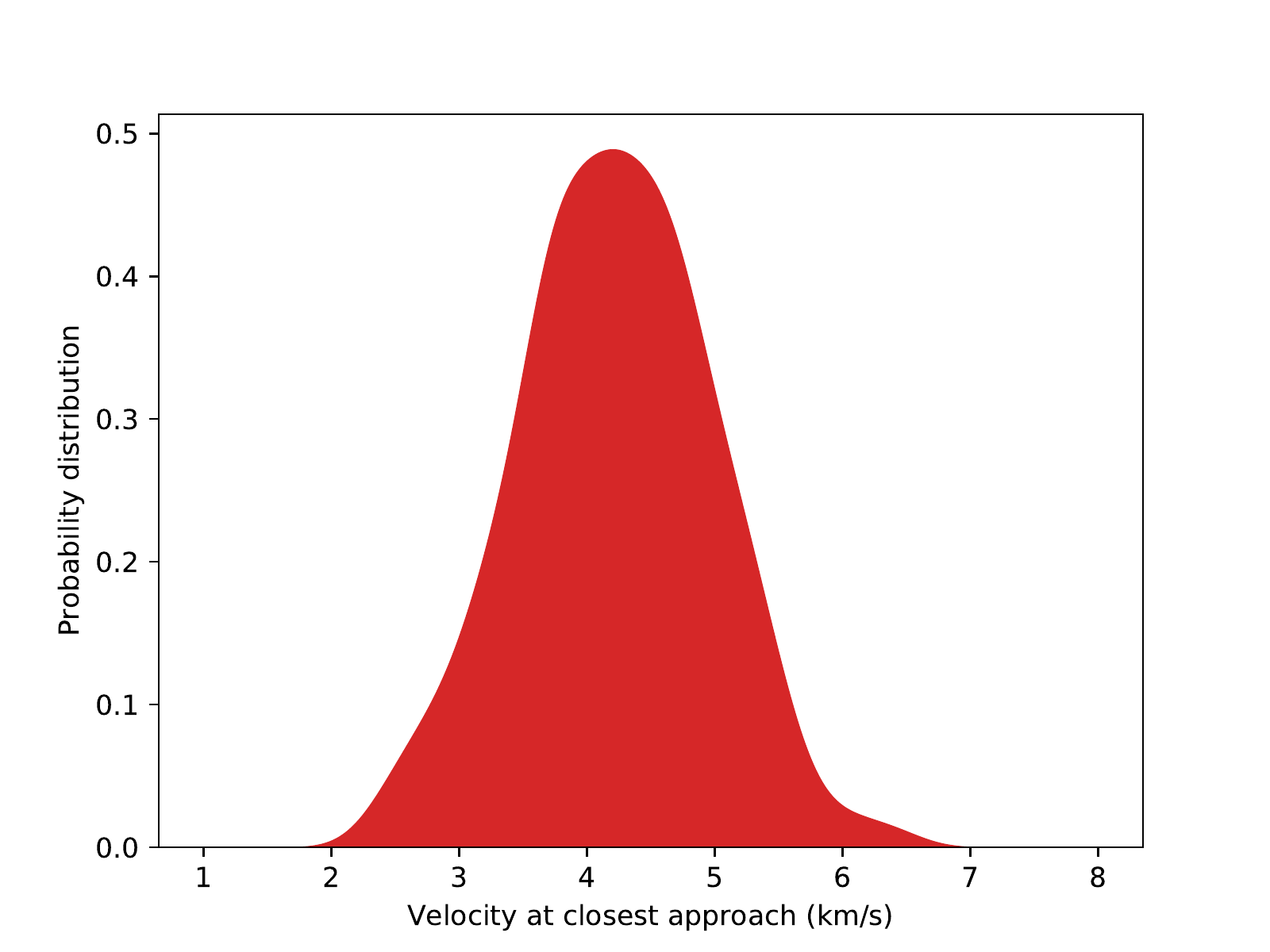}
		\caption{Distribution of the times and velocities at closest approach, for the cases where the distance at closest approach is less than 0.1 pc.}\label{fig:tca}
	\end{figure}
	
	 The distributions of the time and velocities of the perturber at closest approach are represented on Fig. \ref{fig:tca} (only the cases where the distance was less than 0.1 pc). Most of the encounters occur between 4 and 2 Myr ago, with a velocity between 2 and 6 km/s. 
	 
	
	\subsection{Effect on the planet}
	
	\subsubsection{Setup}
	
	Once the configurations for which a close fly-by occur within the 15 Myr of the system life have been identified, we launch a new set of simulations, this time including the planet. The bodies are initialized at their position at the end of the first simulation, that is at their position 15 Myr ago. HD 106906 is separated into two bodies, namely the binary HD 106906 AB ($2.58~\mathrm{M_\odot}$), and the planet HD 106906 ABb ($0.01~\mathrm{M_\odot}$).  The simulations are launched from 15 Myr ago to the present epoch, so that the final outcome represents the current positions of the bodies. The time-step was set to 100 yr, with outputs every 1,000 yr.
	
	In the study of \cite{rodet2017}, the destabilization of the planet takes place after a violent encounter with the central binary, in the beginning of the system's life. The outcome was either a definitive ejection on a hyperbolic trajectory, or a transitional state where the eccentricity raised dramatically without passing $1$. The probability of the different outcomes depends on the characteristics of the encounter, which is highly underconstrained. In the case of a hyperbolic trajectory, a subsequent stabilization by a fly-by must be precisely synchronized, and is thus difficult to achieve. Thus, we study here the case of a highly eccentric transitional bound orbit. The periastron should roughly correspond to the separation of the planet when the perturbation occurred, around 1 au. On the other hand, the apoastron will remained mostly unchanged after a fly-by. The current projected separation implies a minimal value of 730 au. Moreover, the probability is higher to observe the planet near apoastron: it spends 2/3 of its time at a separation greater than 700 au for an apoastron of 1,000 au, and 95 \% for an apoastron of 3,000 au.  All in all, two sets of simulations are performed, where the planet is initialized with a periastron of 1 au and an apoastron of 1,000 ($a = 500.5$ au, $e = 0.998$) or 3,000 au ($a = 1500.5$ au, $e = 0.9993$). 
	
	The necessary energy to completely eject the planet is $\frac{1}{2} GM_\text{HD106906}/a_p$, where $a_p$ is the initial semi-major axis of the planet and $M_\text{HD106906}$ the mass of the host binary. From its current position close to the central binary, a definitive ejection requires around $1~\mathrm{M_\odot au^2/yr^2}$. A proportion of $2.10^{-3}$ less corresponds to an elliptic trajectory with apoastron 1,000 au, and $2.10^{-4}~\mathrm{M_\odot au^2/yr^2}$ less corresponds to 10,000 au. Thus, from an energetic point of view, reaching a high apoastron on a still bound orbit in the ejection process is nearly as costly as being definitely ejected. 
	
	
	
	For a fly-by to have a meaningful role in the dynamical history of the planet, it has to decrease the planet eccentricity by increasing the periastron to a safer value (an increase of the order of the astronomical unit at least). The time-scale of the fly-by is much larger than the orbital period of the planet, so that the initial position of the planet on its orbit is not a relevant parameter in the simulations. Moreover, in our scenario, the planet formed within the disk, so that its orbit was initially coplanar with the disk mid-plane. We assume that the planet apoastron is aligned with the observed extension of the disk. A close encounter with the central binary will retain this coplanarity if the inclination of the binary orbit is similar to that of the disk plane, which seems likely from the first estimates of its orbital parameters \citep{lagrange2019}. As the fly-by is likely to keep the apoastron roughly unchanged and the eccentricity high \citep[consistent with the observed patterns of the disk according to][]{jilkova2015,nesvold2017,rodet2017}, this is consistent with the current position of the planet.
	
	
	\subsubsection{Results}
	
		The conclusion of the study depends essentially on the possibility for the fly-by to increase significantly the periastron. This effect is stongly correlated to the distance at closest approach. We thus represented the periastron change with respect to the distance at closest approach for the outputs of the two sets of simulations on Figs. \ref{fig:periastron1000} and \ref{fig:periastron3000}.
	
		Whether for a 1,000 or 3,000 au apoastron, a 0.1 pc encounter is not enough to significantly raise the periastron: a closer fly-by is required. For the 1,000 au apoastron case, the distance at closest approach must be less than $0.01$ pc, that is $2,000$ au.  For the 3,000 au apoastron case, the destabilization is certainly easier, but the distance at closest approach must still be less than $0.05$ pc, that is $10,000$ au. For such distances, the results are essentially identical for the bound cases, as the separation between the two perturbers is greater of similar than the distance at closest approach with HD 106906. On our initial 10,000 draws, respectively 2 and 20 resulted in a periastron increase superior to 1 au for the 1,000 and 3,000 au apoastron cases, and 1 and 2 lead to the ejection of the planet (for distance at closest approach similar or less than the planet semi-major axis).
	
	Moreover, coplanarity of the planet orbit with the disk plan is expected if the planet formed within the disk. The current projected planet misalignment with the disk plane is currently estimated at 23 degrees, although a lower angle (and even coplanarity) would be possible if the planet true separation is greater than its projected separation ($\gtrsim 3000$ au for coplanarity). A 23\degree~misalignment corresponds to a minimal altitude of $\sim$ 280 au above the disk plane, and such gain of altitude is rarely seen in the simulations, even in the most favorable case of a high initial apoastron. This would suggest that the misalignment (or part of it at least) is an illusion due to projection effects.
	
	\subsubsection{Theory}
	
	We first study the periastron increase as a function of the distance at closest approach, and compare it to the theoretical predictions. The computation of the following theoretical formula is explained in the appendix. The simplest approach is the impulse approximation, where the fly-by is assumed to be instantaneous and trigger a sudden velocity change on the planet. Although this cannot be considered as representative for the reality if we compare the fly-by time-scale with the orbital period of the planet, this approximation often provides a good estimate. In this framework, \cite{brunini1996} show that the fly-by increases the planet velocity by:

	\begin{equation}
		|\Delta v_p| \lesssim \frac{2GM_*}{VD^2} a_p \label{eq:deltav}
	\end{equation}
	
	 \noindent where $v_p$ is the planet velocity, $M_*$ is the perturber's mass, $V$ its velocity at closest approach, $D$ its distance at closest approach, and $a_p$ the planet semi-major axis. This formula nevertheless applies to circular orbits only \citep{brunini1996}. By supposing that the new orbit intersects the old one at apoastron, the planet eccentricity $e_p$ takes part, and we have a change of semi-major axis $\Delta a_p = -a_p \Delta e_p$, which gives a change of periastron $\Delta \text{peri} = - 2 a_p \Delta e_p$. Finally, one gets (see appendix):
	
	\begin{equation}
		|\Delta \text{peri}| \lesssim 8\frac{GM_*}{\sqrt{GM_\text{HD106906}}} \frac{a_p^\frac{5}{2}}{VD^2}
	\end{equation}		 
	 
	 It can be adapted to an eccentric orbit, as was done in \cite{rodet2017}, by supposing that the perturbations occur only at apoastron. Then, stating that the apoastron is preserved, one gets $\Delta a_p = -a_p \Delta e_p/(1+e_p)$ and $\Delta \text{peri} = - 2 a_p \Delta e_p/(1+e_p)$. Finally, using Eq. \ref{eq:deltav} to quantify the velocity increase at apoastron, one gets (see appendix):
	   
	\begin{equation}
		|\Delta \text{peri}| \lesssim 8\frac{GM_*}{\sqrt{GM_\text{HD106906}}} \frac{a_p^\frac{5}{2}}{VD^2} \frac{\sqrt{(1-ep)(1+e_p)}}{3-e_p}\label{eq:apoimpulse}
	\end{equation}
	
	On the other hand, a more rigorous approach is to compute the secular evolution of the orbital elements of the planet during the passage of the perturber. \cite{heggie1996} used that method to determine the eccentricity increase of a companion, and found a complex formula depending on all 6 orbital elements of the perturber's orbit. In this framework, the semi-major axis is invariant throughout the fly-by. Considering a coplanar orbit and a perturber's eccentricity significantly higher than 1 (strongly unbound orbit), the maximum is:
	
	\begin{equation}
		|\Delta \text{peri}| \lesssim \frac{5}{2} \frac{GM_*}{\sqrt{GM_\text{HD106906}}} \frac{a_p^\frac{5}{2}}{VD^2} e_p \sqrt{1-ep^2} \label{eq:secular}
	\end{equation}
	
%
	
\begin{figure}
		\includegraphics[width=\linewidth]{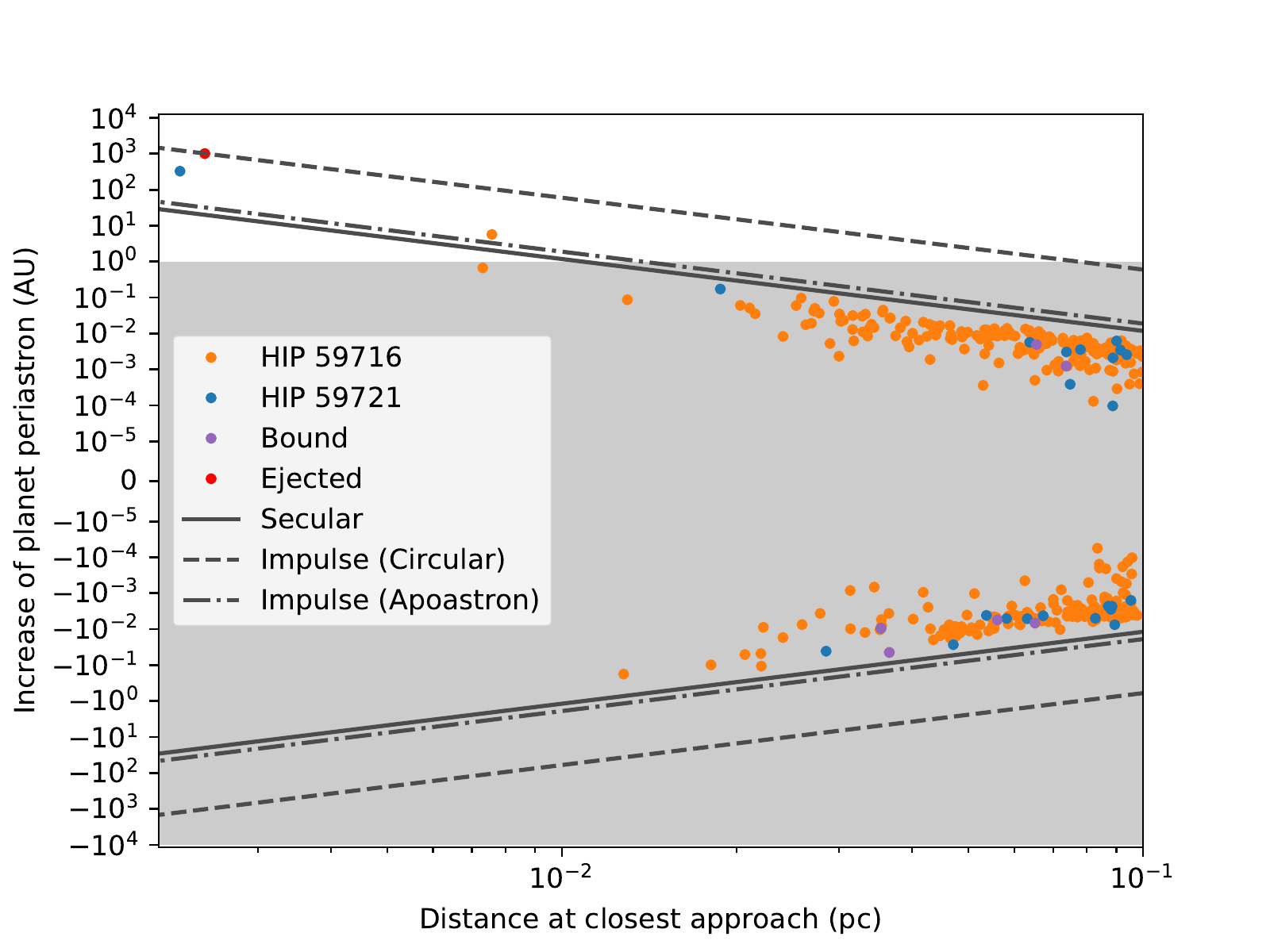}
		\caption{Periastron increase with respect to the distance at closest approach, from N-body simulations (dots) and theoretical approaches (lines), for the closer fly-bys, and for an initial planetary apoastron of 1,000 au. The grey part corresponds to a periastron change inferior to $+1$ au, which will not secure the planet stability.}\label{fig:periastron1000}
	\end{figure}

\begin{figure}
		\includegraphics[width=\linewidth]{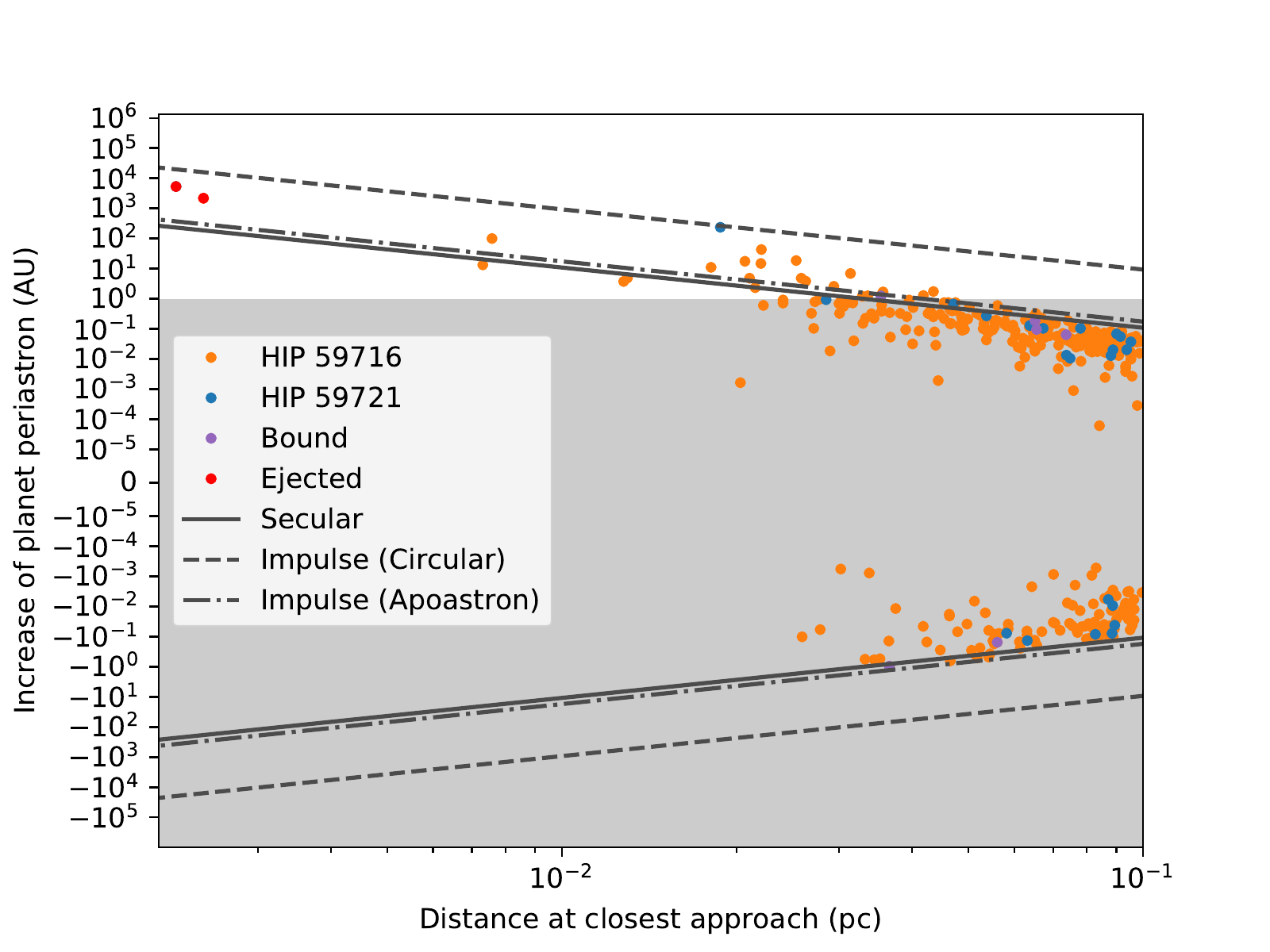}
		\caption{Periastron increase with respect to the distance at closest approach, from N-body simulations (dots) and theoretical approaches (lines), for the closer fly-bys, and for an initial planetary apoastron of 3,000 au. The grey part corresponds to a periastron change inferior to $+1$ au, which will not secure the planet stability.}\label{fig:periastron3000}
	\end{figure}

	The three theoretical predictions are represented on Figs. \ref{fig:periastron1000} and \ref{fig:periastron3000}: circular impulse, apoastron impulse and secular approximation. They all correspond to maximum values, as the true periastron evolution depends on the angular characteristics of the encounter. The velocity $V$ is set to its mean value over all closest approaches, around 4 km/s. $M_*$ was set to $1.3~\mathrm{M_\odot}$, but the increase depends weakly on the perturber's exact mass. The eccentricity $e_p$ is set to its initial value, an approximation that becomes less relevant when $\Delta e_p \gtrsim 1 - e_p = 2. 10^{-3}$ (for closest approach less or around 0.01 pc).
	
	We see on Fig. \ref{fig:periastron1000} that the periastron change is best modeled by the secular approximation, but is also correctly approached by the impulse approximation at apoastron. It suggests that the effect of both perturbers on the planet can be estimated by the effect of the perturber that had the closest approach. This is also true for the cases where the two perturbers are bound (see Appendix).
		
	Furthermore, we seek to estimate if the fly-by could account for the possible misalignment of the planet with the debris disk plane. Depending on the exact value of the argument of periastron $\omega_p$, a very eccentric orbit does not necessarily have a large elevation above the disk plane, even if it is highly inclined. To have a meaningful plan misalignment, the planet should have an inclination change combined with a shift of the argument of its periastron that results in a significant elevation above the disk plane. For any Keplerian orbit, the maximum elevation $z_\text{max}$ above the reference plane is given by:
	
		\begin{equation}
			z_\text{max} = a_p \sin(i_p) \left(\sqrt{1-e_p^2\cos^2(\omega_p)}+e_p |\sin(\omega_p)| \right) \quad .
		\end{equation}
		
		\noindent Obviously, with $e_p \sim 1$ and $\omega_p \sim 0$ or $\pi$, $z_\text{max}$ remains small irrespective of the value of $i_p$.
	
\begin{figure}
		\includegraphics[width=\linewidth]{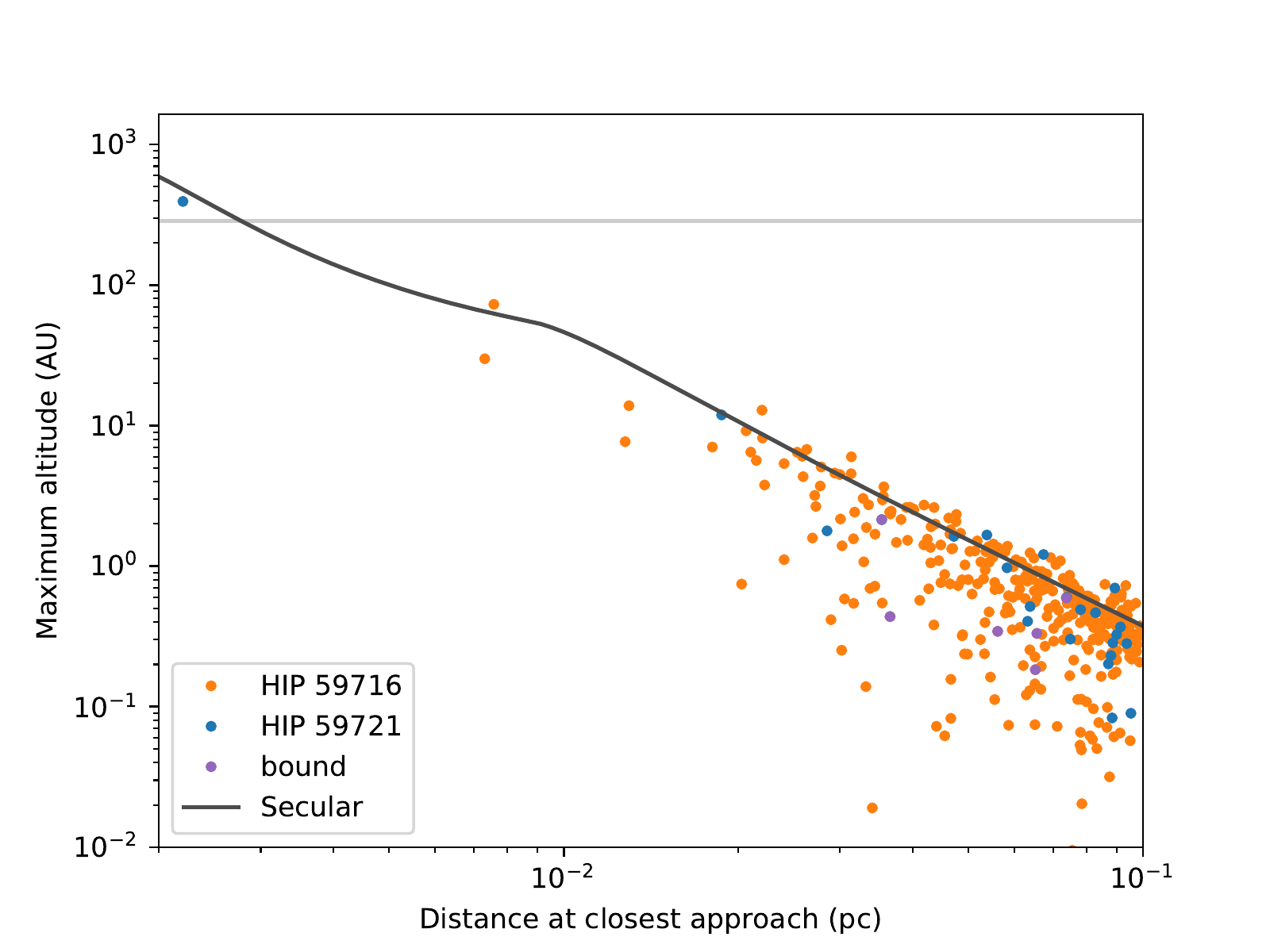}
		\caption{Maximal altitude with respect to the distance at closest approach, from N-body simulations (dots) and secular theoretical approach (line), for the closer fly-bys, and for an initial planetary apoastron of 1,000 au. The grey line indicates the projected elevation of the planet.}\label{fig:zmax1000}
	\end{figure}

\begin{figure}
		\includegraphics[width=\linewidth]{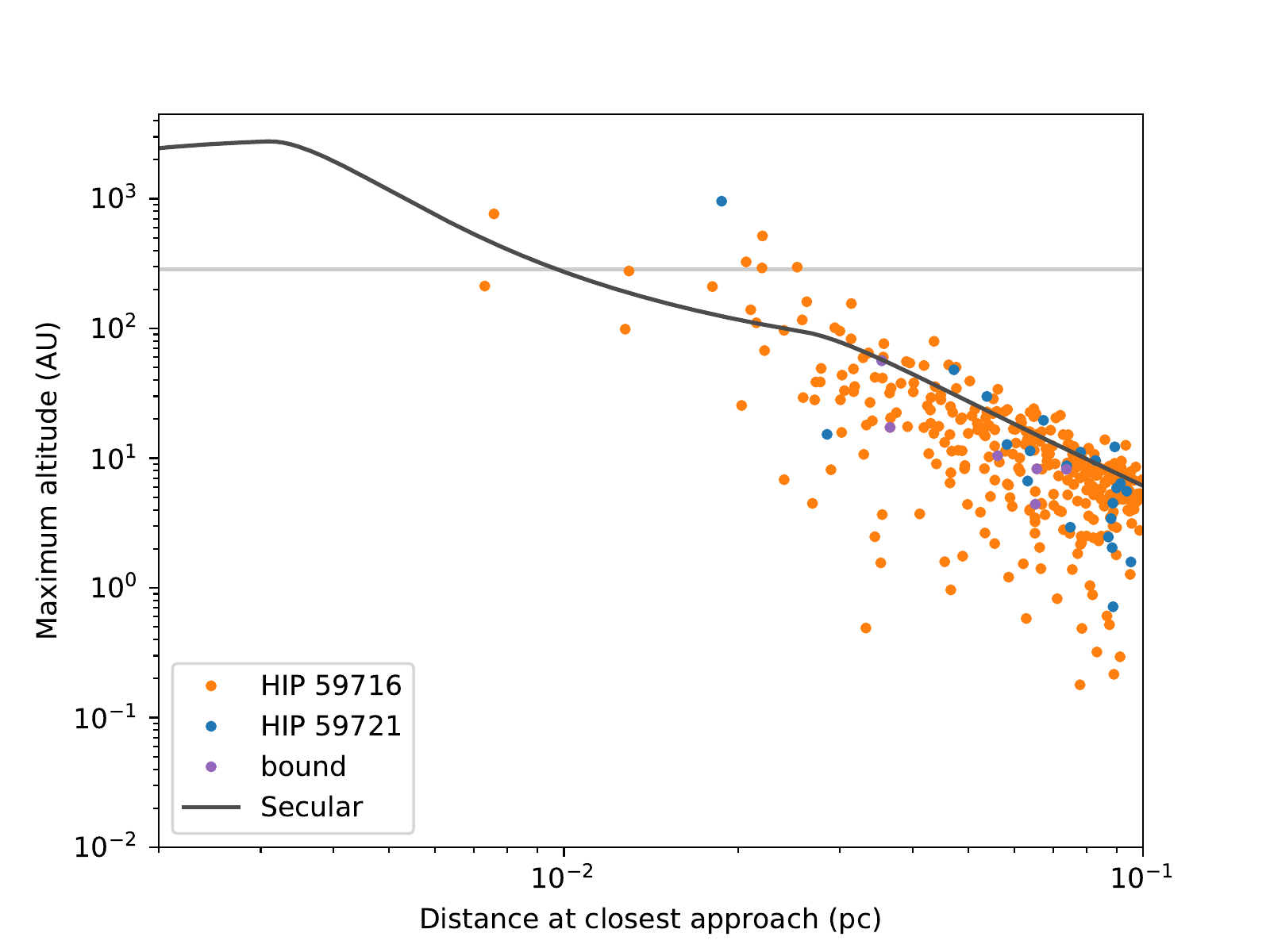}
		\caption{Maximal altitude with respect to the distance at closest approach, from N-body simulations (dots) and secular theoretical approach (line), for the closer fly-bys, and for an initial planetary apoastron of 3,000 au. The grey line indicates the projected elevation of the planet.}\label{fig:zmax3000}
	\end{figure}

	We thus computed the change in $z_\text{max} $, inspiring from \cite{heggie1996}. The details are explained in the appendix. The resulting maximal altitude is represented on Fig.\ref{fig:zmax1000} and \ref{fig:zmax3000}.
	
	\subsubsection{Discussion}	
	
	From both approaches, theoretical and numerical, in the most favorable case, it appears that a fly-by has a significant impact on the planet (periastron increase above $1$ au) only if its closest approach is less than 0.05 pc, that is 10,000 au. This corresponds to a small subset among the initial draws, not because of an incompatibility with the observations, but because of the high dispersion of closest approaches, underconstrained by the observations.

	
	We checked that the distance at closest approach is not correlated to the time at closest approach, nor to the velocity at closest approach. Considering the compatibility between our results and the dynamical scenario proposed in \cite{rodet2017}, the time of the fly-by must be considered. Given our simulations, the closest approach occurred likely 2 to 4 Myr ago  ($3 \pm 1$ Myr). However, our scenario account for the ejection of the planet only in the beginning of the system life, when protoplanetary disk is still present and can effectively trigger planetary migration. Given the disk lifetime for massive stars \citep[$\sim 3$ Myr,][]{ribas2015} and the system assumed age (15 Myr), 2 to 4 Myr ago is significantly too late for the fly-by to have a decisive role. However, a younger age for the system (10 Myr, compatible with LCC age spread of 6 Myr) could still account for this discrepancy.
	
	\subsection{Effect on the disk}
	
	The effects of a fly-by on a disk may be significant, depending on the parameters of the encounter. The case of a dynamically efficient fly-by can be observed in system  HD 141569, where the ongoing encounter has been deeply studied in \cite{reche2009}. In this system, the fly-by could be responsible for truncation, spiral formation, collisional evolution, eccentricity and inclination raise. In our study, the effect of the fly-by on test-particles will be essentially similar to that on the planet. Since the test particles in a debris disk have a nearly circular orbit, the fly-by will increase the eccentricity, significantly or not depending on the distance of closest approach. Moreover, all fly-by characteristics being equal, particles inclination will be excited differently depending on their distance to the host star. The disk might then be warped. The sensitivity of the scattered-light images of the disk are not sufficient to reveal a weak warp, but the warp can induce further instabilities and asymmetries in the disk that could account for its non-standard shape.
		
\begin{figure}
		\centering
		\includegraphics[width=\linewidth]{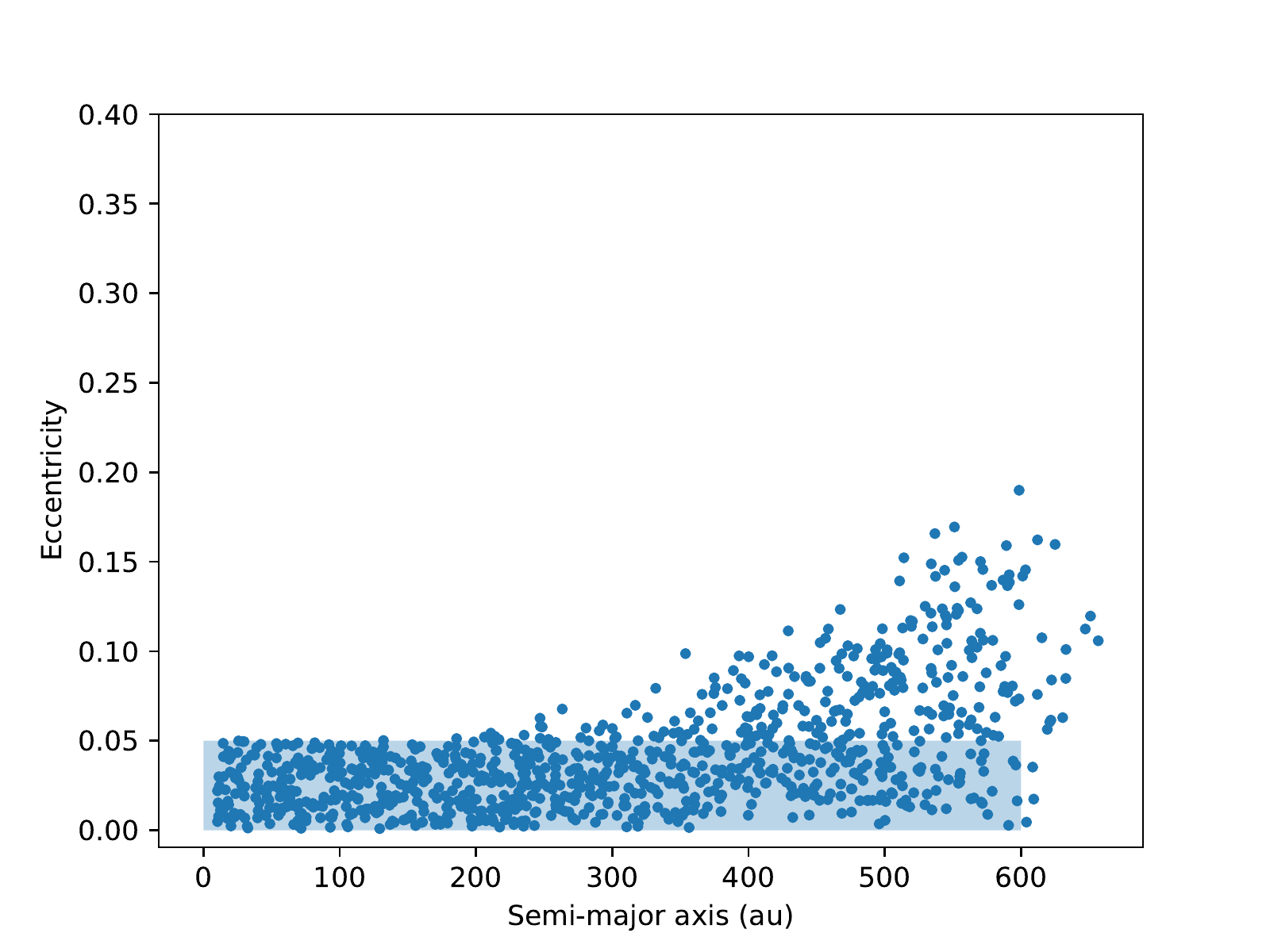}
		\includegraphics[width=\linewidth]{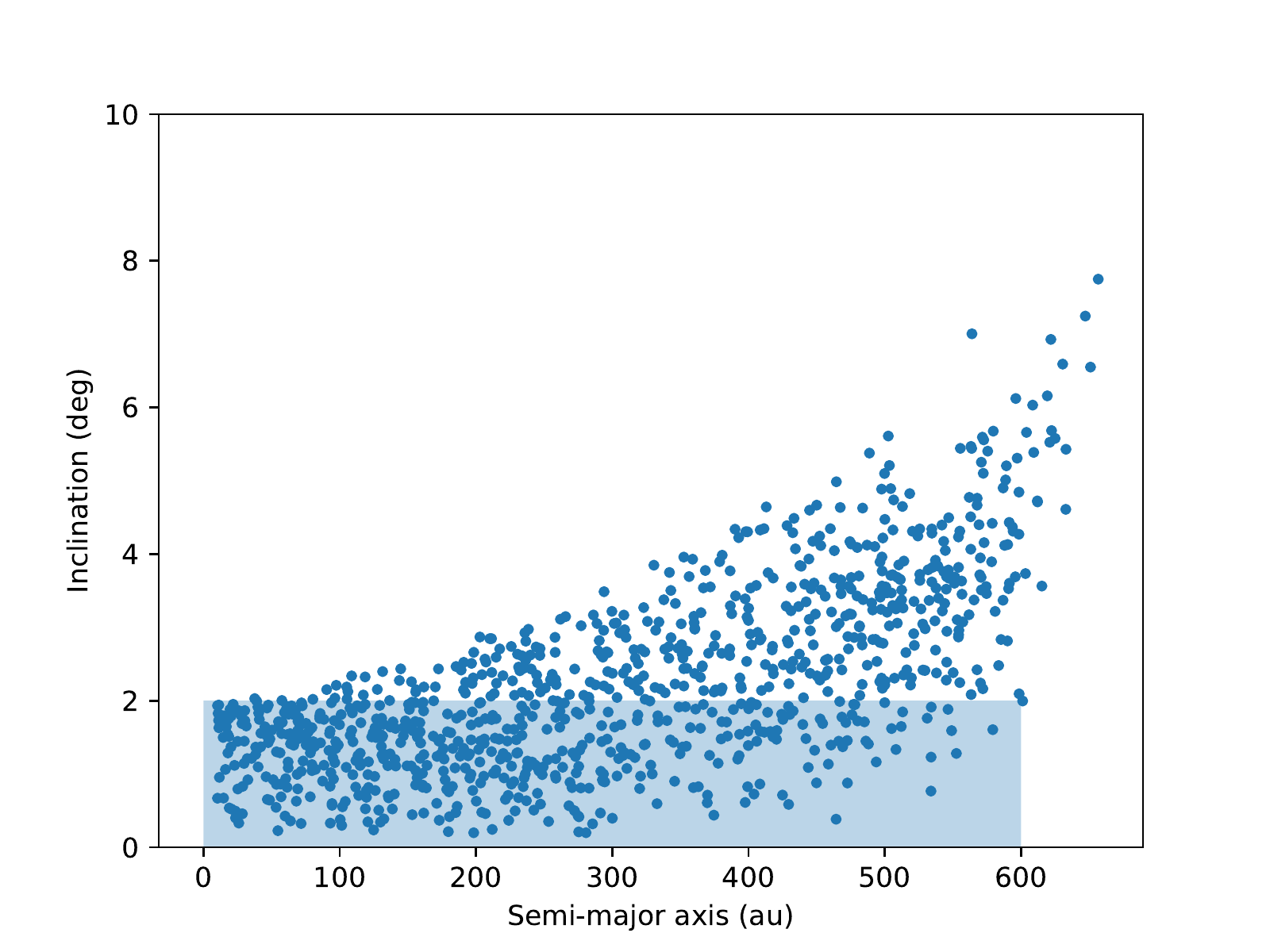}
		\caption{Orbital elements of the test particles after a close fly-by. The lightly blue zones represent the initial configuration.}\label{fig:disk}
	\end{figure}

		We chose among the previous cases a situation with a very short distance at closest approach (1,000 au), with a medium relative inclination ($\sim$ 45 \degree) and ran a simulation with the three massive bodies (HD 106906 ABb and the perturbers) and 1,000 test particles. The particles have initially semi-major axes evenly shared between 10 and 600 au, eccentricity below 0.05, and an inclination spread of 2 degrees. The simulation was launched for 100 000 years around the fly-by epoch, with a time step of 1 yr. The resulting disk is represented on Fig. \ref{fig:disk}.
		
		On the other hand, the repeating passing of the planet within the disk would have stronger consequences. If a very small percentage is ejected over one period ($\lesssim 0.01$ \%), the mean eccentricity of the particles raise from 0.02 at each passage. For the disk to remain long-lived in its current shape, \cite{jilkova2015} (non collisional simulations) and \cite{nesvold2017} (collisional simulations) estimated that the planet orbit should not cross the disk. Thus, the planet periastron should be outer to the observed $\sim$ 100 au outer disk radius. Within our scenario, it means that this enlargement of the periastron occurred rather quickly, whether or not it was caused entirely by the fly-by. In any case, the planet interactions would have cover the track of the fly-by-induced perturbations
		
		The new structure of the code allows to estimate the percentage of dust capture by the planet. It turns out that temporary (less than 10 yr) capture is experienced by about 5\% of the dust at each passage, but no permanent captures were produced.
	
	\section{Conclusion}
	
	In this paper, we present the N-body mixed-variable code \texttt{ODEA}, that is able to study multiple systems in evolving architectures. We use it to study the rare planetary system HD 106906. We confirm that the two stars identified by \cite{derosa2019} could have helped stabilizing the planet after a destabilization by its host binary star. This scenario could account for the wide separation of the planet, its possible elevation with respect to the disk plane, as well as the structures evidenced within the disk.
	
	However, the significance of the encounter strongly depends on the distances at closest approach. With the current precision on the three systems configuration (especially the relative radial velocities and distances), it is not possible to establish the role of the flybys. To circularize the planet orbit if it was previously ejected on a wide trajectory, a fly-by closer than 0.05 pc is needed (assuming apoastron $\leq$ 3,000 au), which is one order of magnitude below the uncertainty on the closest approach. The simulations show that the angular configuration is favorable when this condition is met. 
	
	
	Any indication of HD 106906 b relative motion would be helpful to constrain its orbit, and thus its dynamical history. More precise parallaxes and radial velocities for HIP 59716 and HIP 59721 are necessary to constrain the distances at closest approach, and conclude on the effect of the fly-bys on the system dynamical evolution.
	
	\texttt{ODEA} handles hierarchy changes in systems with non-Solar-system-type architectures. It can model efficiently captures and fly-bys. Through a criterion based on accelerations ratios, a new hierarchy is defined when the current is perturbed. \texttt{ODEA}'s natural upgrade is the implementation of a Mercury-like approach to handle close encounters, that is transitional states of non-Keplerian movements. 
	
	\begin{acknowledgements}
		We thank the anonymous referee for reviewing our work and for insightful and constructive comments which improved the manuscript. The   project   is   supported   by   CNRS,   by   the   Agence Nationale de la Recherche (ANR-14-CE33-0018, GIPSE), the OSUG@2020 labex and the Programme National de  Planétologie  (PNP,  INSU)  and  Programme  National  de  Physique  Stellaire (PNPS,  INSU). Most of the computations presented in this paper were performed using the Froggy platform of the CIMENT infrastructure (https://ciment.ujf-grenoble.fr), which is supported by the Rh\^one-Alpes region (GRANT CPER07\_13 CIRA), the OSUG@2020 labex (reference ANR10 LABX56) and the Equip@Meso project (reference ANR-10-EQPX-29-01) of the programme Investissements d'Avenir, supervised by the Agence Nationale pour la Recherche.  P.K. and R.J.D.R. thank support from NSF AST-1518332, NASA NNX15AC89G and NNX15AD95G/NEXSS. This work benefited from NASA’s Nexus for Exoplanet System Science (NExSS) research coordination network sponsored by NASA's Science Mission Directorate.
	\end{acknowledgements}
	
	\bibliographystyle{aa}
	\bibliography{../Biblio}
	
	\section*{Appendix}
	
	\subsection*{Test simulation for \texttt{ODEA}}
	
		The simulation that was used for the study of the performance of  \texttt{ODEA} belongs to the 10,000 simulations performed in section \ref{sec:simulations}, in the case where the two perturbers were bound. We chose a simulation with a small distance of closest approach ($\sim$ 2,000 au), so that there is an effect on the planet orbital elements. The fly-bys occurred between 4 and 2 Myr ago, so that we restricted ourselves to this time interval when studying the energy error with respect to the time-scale (to limit the floating point round-off error associated with the large distances).
		
	\subsection*{Theoretical error associated with the symplectic mapping}
		
		Splitting the Hamiltonian with a kick-drift-kick approach, as described in Sec. \ref{sec:structure}, the energy error that we get is \citep{saha1994}
	
		 \begin{equation}
			\tilde{H} = H - \frac{\Delta t^2}{12} \lbrace \lbrace \ha, \hb \rbrace , \ha + \frac{1}{2} \hb \rbrace + O(\Delta t ^4) \quad,
		\end{equation}
		
		\noindent where the Poisson brackets are defined as follows:

		\begin{equation}
			\lbrace f, g \rbrace = \sum_i \frac{\partial{f}}{\partial{\vec{r_i}}} \frac{\partial{g}}{\partial\vec{p_i}} - \frac{\partial{f}}{\partial\vec{p_i}} \frac{\partial{g}}{\partial\vec{r_i}}\quad.
		\end{equation}
			
		In \texttt{SWIFT HJS}, $\ha$ and $\hb$ are given by equations \ref{eq:ha} and \ref{eq:hb}, which can be computed respectively with $O(N)$ and $O(N^2)$ operations. Thus, computing $\lbrace \lbrace \ha, \hb \rbrace, \ha \rbrace$ already requires $O(N^4)$ operations.
		
%

	\subsection*{Derivation of the changes of planet periastron due to the fly-by in the impulse approximation}
	
		\subsubsection*{Circular impulse}
		
		The expression of the change of the planet velocity is given in Eq. \ref{eq:deltav}. Supposing that the new orbit intersects the old one at apoastron or periastron, we have $\Delta a_p = a_p \Delta e_p$. Moreover, the velocity of the planet if on a circular orbit is $v_p = \sqrt{{GM_\text{HD106906}}/{a_p}}$. Thus, the eccentricity is 
		
		\begin{equation}
			|\Delta e_p| = \frac{|\Delta a_p|}{a_p}  = 2\frac{|\Delta v_p|}{v_p} \lesssim 4\frac{GM_*}{\sqrt{GM_\text{HD106906}}} \frac{a_p^\frac{3}{2}}{VD^2}\nonumber
		\end{equation}	
		
		\noindent and the periastron is then given by $\Delta \text{peri} = \Delta a_p - a \Delta e_p = - 2 a_p \Delta e_p$.
	
		\subsubsection*{Apoastron impulse}	 
	 
		Stating instead that the apoastron is preserved, one gets $\Delta a_p = -a_p \Delta e_p/(1+e_p)$. Within the impulse framework, the change of velocity involves the velocity at apoastron, so that the velocity writes $v_p = \sqrt{{GM_\text{tot}}/{a_p}} \sqrt{(1-e)/(1+e)}$ . Thus,
		
		\begin{equation}
			\frac{\Delta v_p}{v_p} = -\frac{\Delta a_p}{2a_p} - \frac{\Delta e_p}{1-ep^2} = -\frac{\Delta e_p}{2(1+e_p)} - \frac{\Delta e_p}{1-ep^2} \nonumber
		\end{equation}
		
		\noindent which gives
		
		\begin{align*}
			|\Delta e_p| &= -2\frac{|\Delta v_p|}{v_p} \frac{1-e_p^2}{3-e_p} \\
			&\lesssim 4\frac{GM_*}{\sqrt{GM_\text{HD106906}}} \frac{a_p^\frac{3}{2}}{VD^2} \frac{(1+ep)^\frac{3}{2}\sqrt{1-e_p}}{3-e_p} \nonumber
		\end{align*}
		
		\noindent and the periastron is then given by $\Delta \text{peri} = \Delta a_p (1-e_p) - a \Delta e_p = - 2 a_p \Delta e_p/(1+e_p)$.
	
	\subsection*{Derivation of the changes of planet orbital characteristics due to the fly-by in the secular approximation}	
	
		\subsubsection*{Perturbative potential}
		
		We inspire from \cite{heggie1996} to derive the first-order perturbation of the planet orbital elements in the secular approximation. 
		
		Following Heggie \& Rasi, we number respectively $1$, $2$ and $3$ HD 106906 central star, HD 106906 b and one of the stellar perturber. The position of the planet relative to its host star is denoted by $\vec{r}$, and the position of the third body relative to HD 106906 center of mass is denoted by $\vec{R}$. In this framework, the evolution of the planet orbit verifies:
		
		\begin{align*}
			\vec{\ddot{r}} &= -\frac{GM_{12}}{r^3} \vec{r} + \vec{\nabla} U\\
			U &= \frac{G m_3 M_{12}}{m_1 m_2} \left( \frac{m_2}{|\vec{R} - \frac{m_1}{M_{12}}\vec{r}|} -  \frac{m_1}{|\vec{R} + \frac{m_2}{M_{12}}\vec{r}|} \right) \\
			&= \frac{G_m3r^2}{2R^3} \left( 3 (\frac{\vec{r}.\vec{R}}{rR})^2 -1 \right) + O((\frac{r}{R})^3)
		\end{align*}
		
		\noindent where $U$ is the perturbative potential. 
		
		In the secular approximation, $U$ is averaged over the orbit of HD 106906 planetary orbit. The implicit assumptions is that all orbital elements but the anomaly have a longer evolution timescale than the orbital period. As we are interested in the first order evolution, we only integrate the dominant part in $a_p/a$ (quadripole order). Then, we use Lagrange equations to retrieve the evolution of the eccentricity, the inclination and the longitude of periastron.
		
		\subsubsection*{Eccentricity and periastron change}
		
		After we first averaged over the planet orbital motion, the secular evolution of the eccentricity obtained at the quadrupole level writes:
		
		\begin{equation*}
			\frac{de_p}{dt} = \frac{15Gm_3 R_x R_y a_p^\frac{3}{2} e_p \sqrt{1-ep^2}}{2 R^5 \sqrt{GM_{12}}}
		\end{equation*}
		
		\noindent where the x-y plane is the initial plane of the planet (plane of the disk), and the $x$ direction is given by the planet initial periastron. To compute the first order of the change of $e$ after the fly-by, we integrate $de/dt$ along time from $-\infty$ to $+\infty$ by fixing all variables to their initial values but the angular evolution of the stellar perturber.
		
		Heggie \& Rasio computed in their Eq. (7) the change in eccentricity as a function of the angular parameters of the encounter, and we exactly retrieve their expression. The maximum efficiency is obtained for a coplanar encounter, where all the transferred angular momentum apply only on the eccentricity. Stating that the eccentricity of the perturber's orbit is significantly more than $1$ ($V = 3$ km/s and $D = 1$ pc gives $e \sim 500$, $D = 0.1$ pc gives $e \sim 50$), we obtain
		
		\begin{equation}
			\Delta e_p = - \frac{5}{2} \frac{M_*}{\sqrt{M_\text{HD106906}M_\text{tot}}} \frac{a_p^\frac{3}{2}}{D^\frac{3}{2}} \frac{e_p \sqrt{1-e_p^2} }{\sqrt{e}} \sin(2\Omega+2\omega) \nonumber
		\end{equation}
		
		\noindent where $\Omega$ is the longitude of the ascending node and $\omega$ the argument of the periastron of the perturber hyperbolic orbit. The maximum is obtained for $\Omega+\omega = \pi/4$. Moreover, the eccentricity $e$ depends on $D$, $V$ and $GM_\text{tot}$ as $V = \sqrt{GM_\text{tot}(1+e)/D}$ so that $\sqrt{e} \simeq V\sqrt{D/GM_\text{tot}} $. Thus, the eccentricity change satisfies:
		
		\begin{equation}
			|\Delta e_p| \lesssim \frac{5}{2} \frac{GM_*}{\sqrt{GM_\text{HD106906}}} \frac{a_p^\frac{3}{2}}{VD^2} e_p \sqrt{1-ep^2}
		\end{equation}
		
		On the other hand, the semi-major axis is constant in the secular approximation. The periastron is then given by $\Delta \text{peri} = -a \Delta e_p$.
		
		\subsubsection*{Inclination change}
		
		The secular evolution of the inclination obtained at the quadrupole level writes:
		
		\begin{equation*}
			\frac{di_p}{dt} = -\frac{3  G m_3 a_p^\frac{3}{2} \left(4 e_p^2+1\right) R_x R_z}{2 R^5 \sqrt{\left(1-e_p^2\right) G M_{12}}}
		\end{equation*}
		
		We then integrate as before to compute the change of inclination $\Delta i_p$.
		
		\begin{align*}
			\Delta i_p = &\frac{3}{2} \frac{GM_*}{\sqrt{GM_\text{HD106906}}} \frac{a_p^\frac{3}{2}}{VD^2} \frac{1+4 e_p^2}{\sqrt{1-e_p^2}}\\
		& \left(\cos(i)\sin(\Omega) (\arccos(-\frac{1}{e})+\sqrt{e^2-1})  \right.\\
		& \left. - (\cos(\Omega)\sin(2\omega)+\cos(i)\sin(\Omega)\cos(2\omega))\frac{(e^2-1)^\frac{3}{2}}{3e^2} \right)
		\end{align*}
		
		The maximum is reached for $i = \pi/4$, $\Omega = \pi/2$ and $\omega = \pi/2$. Thus, we obtain
		
		\begin{equation*}
			\Delta i_p \lesssim \frac{GM_*}{\sqrt{GM_\text{HD106906}}} \frac{a_p^\frac{3}{2}}{VD^2} \frac{1+4 e_p^2}{\sqrt{1-e_p^2}}
		\end{equation*}
		
		\subsubsection*{Longitude of the periastron change}
		
		The secular evolution of the total longitude of the periastron $\bar{\omega_p} = \omega_p+\Omega_p$ obtained at the quadrupole level writes:
		
		\begin{equation*}
			\frac{d\bar{\omega_p}}{dt} = -\frac{3  G m_3 a_p^\frac{3}{2} \sqrt{\left(1-e_p^2\right)} (R^2 - 4R_x^2 + R_y^2)}{2 R^5 \sqrt{ G M_{12}}}
		\end{equation*}
		
		We then integrate as before to compute the change of inclination $\Delta\bar{\omega_p} _p$.
		
		\begin{align*}
			\Delta \bar{\omega_p}  = &\frac{1}{4} \frac{GM_*}{\sqrt{GM_\text{HD106906}}} \frac{a_p^\frac{3}{2}}{VD^2} \sqrt{1-e_p^2}\\
		& \left( 6\cos^2(i)\cos^2(\omega) - 5(\cos(2i)-3)\cos^2(\omega)\cos(2\Omega) \right.\\
		& \left. +2\cos(2i) (3-5\cos(2\Omega))\sin^2(\omega) - 10\cos(i)\sin(2\omega)\sin(2\Omega) \right)
		\end{align*}
		
		The maximum is reached for $i = \pi/2$, $\Omega = 0$ and $\omega = 0$. Thus, we obtain
		
		\begin{equation*}
			\Delta \bar{\omega_p} \lesssim 5 \frac{GM_*}{\sqrt{GM_\text{HD106906}}} \frac{a_p^\frac{3}{2}}{VD^2} \sqrt{1-e_p^2}
		\end{equation*}
		
		\subsubsection*{Maximal altitude}
		
		The maximum altitude $z_\text{max}$ reached by the planet on its orbit is given as a function of its orbital elements:
		
		\begin{equation}
			z_\text{max} = a_p \sin(i_p) \left(\sqrt{1-e_p^2\cos^2(\omega_p)}+e_p |\sin(\omega_p)| \right) \quad .
		\end{equation}
		
		\noindent It thus depends on the evolution of $a_p$, $e_p$, $i_p$ and $\omega_p$. 
		
		 Due to the term $\sin(i_p)$, the same approach than above leads to neglecting all evolution but that of the inclination. It is consistent with the fact that in the previous expressions, $\Delta_ip \gg \Delta e_p, \Delta i_p$ when the eccentricity tends to $1$. We get:
		 
		 \begin{align}
		 	\Delta z_\text{max} &= a_p \sqrt{1-e_p^2} \Delta i_p\\
		 		&\lesssim \frac{GM_*}{\sqrt{GM_\text{HD106906}}} \frac{a_p^\frac{5}{2}}{VD^2} (1+4 e_p^2) \quad .
		 \end{align}
		
		However, this estimate is not valid anymore when $\Delta_ip$ approaches $\pi/2$, that is when $\sin(i_p)$ approaches $1$. At this point, the estimates of $\Delta e_p$ and $\Delta \bar{\omega}$ must be taken into account. In order to comprise all the different evolution scales, we thus simply estimate the maximal altitude by replacing directly the computed evolution in the definition formula:
		
		\begin{equation}
			\Delta z_\text{max} \lesssim a_p \sin\left(\tilde{i_p}\right) \left(\sqrt{1-\tilde{e_p}^2 \cos^2\left(\tilde{\omega}\right)} + \tilde{e_p}|\sin \left(\tilde{\omega}\right) | \right)
		\end{equation}
		
		\noindent where $\tilde{i_p} = \max(\Delta i_p,\frac{\pi}{2})$, $\tilde{e_p} = e_p-\Delta e_p$ and $\tilde{\omega} = \max(\Delta\bar{\omega}_p,\frac{\pi}{2})$.
		
		 	\subsection*{Additional materials for HD 106906 fly-by simulations}
 	
\begin{figure*}[ht]
		\includegraphics[width=\linewidth]{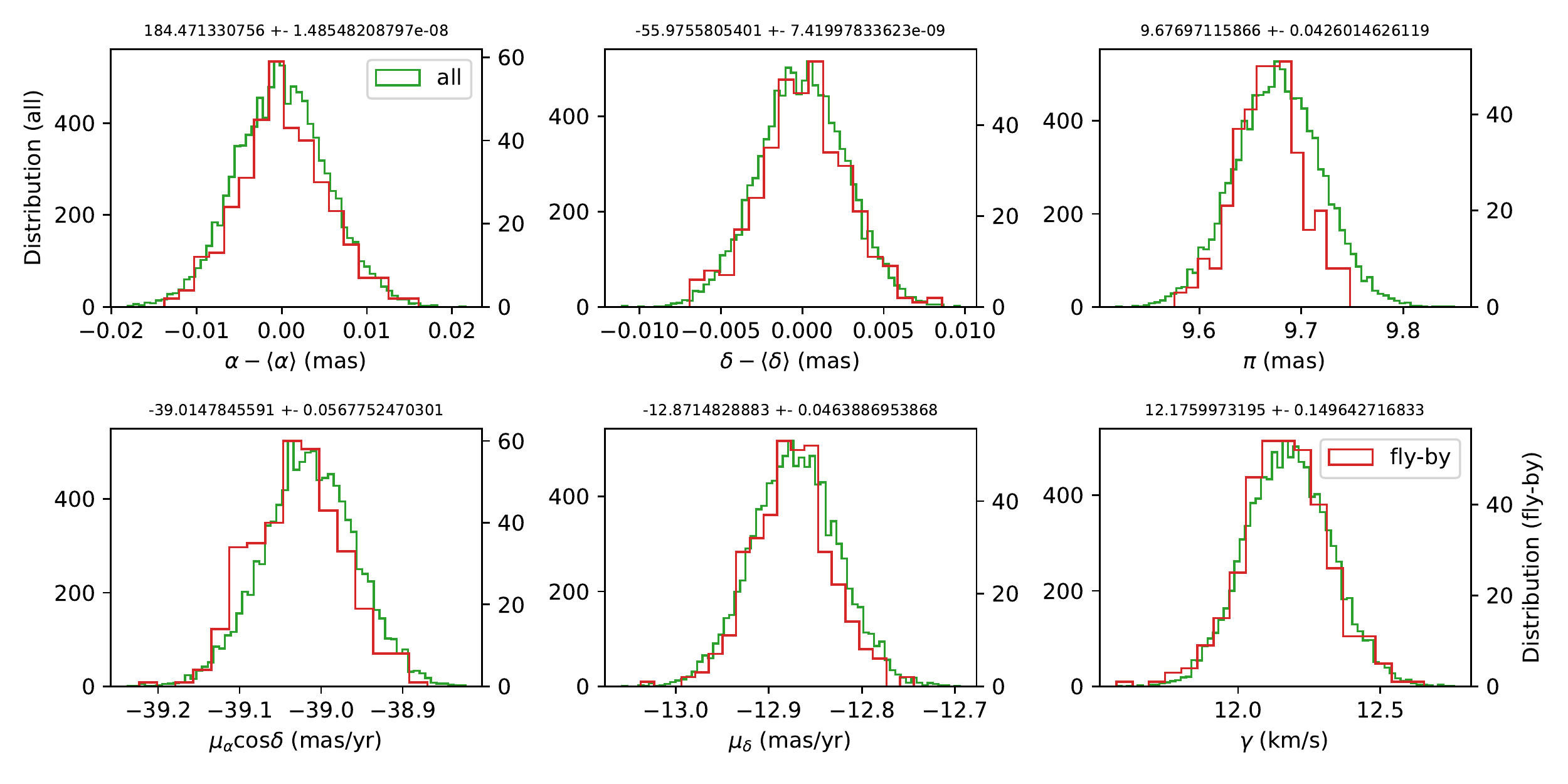}
		\caption{Initial distribution (today) of HD 106906 coordinates and velocities for the 10,000 simulated cases (green), and for the 359 cases where a fly-by closer than 0.1 pc occurred (red).}\label{fig:coordinates1}
	\end{figure*}

\begin{figure*}
		\includegraphics[width=\linewidth]{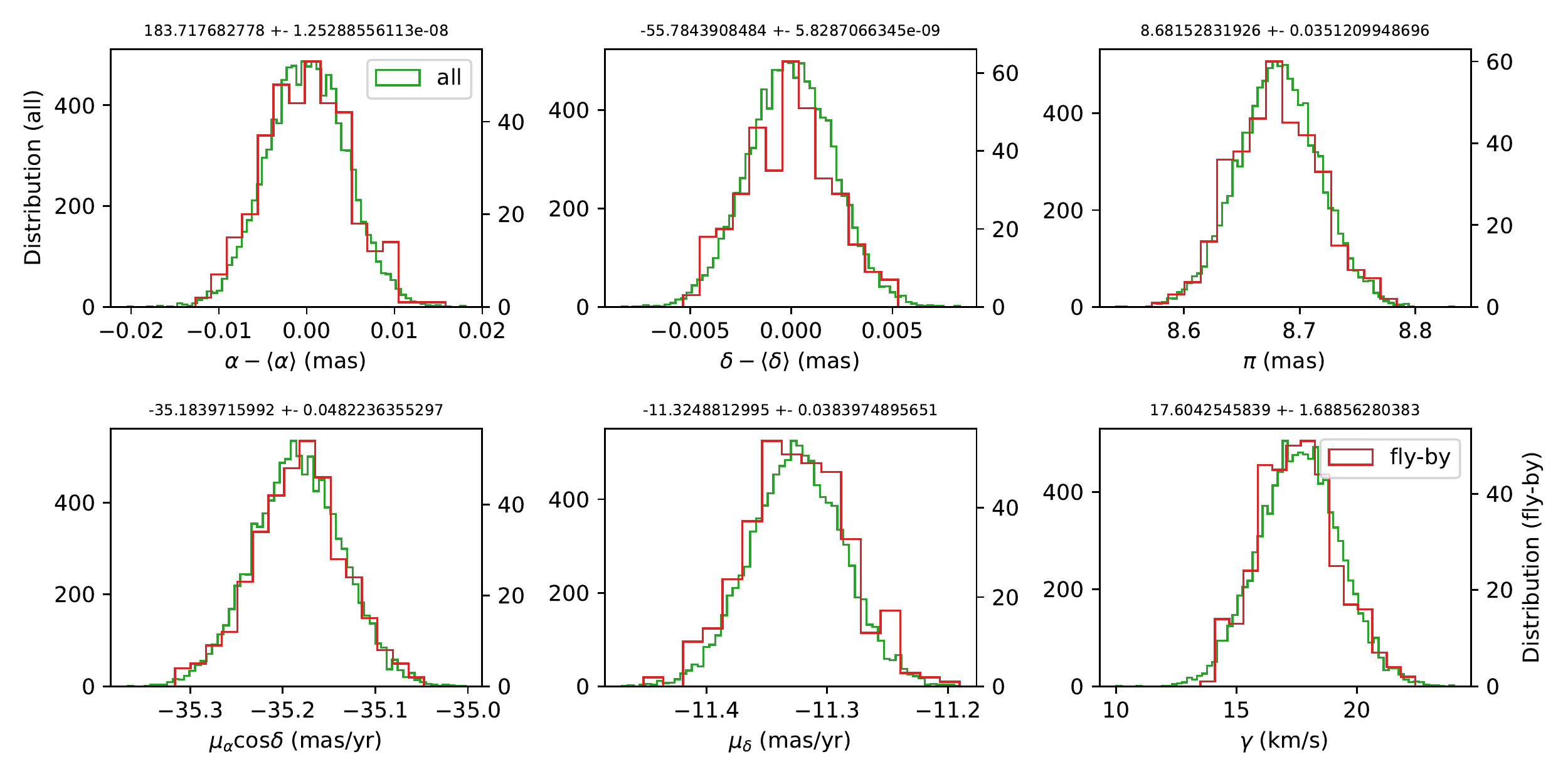}
		\caption{Initial distribution (today) of HIP 59721 coordinates and velocities for the 10,000 simulated cases (green), and for the 359 cases where a fly-by closer than 0.1 pc occurred (red).}\label{fig:coordinates3}
	\end{figure*}

 	Fig. \ref{fig:coordinates1} and \ref{fig:coordinates3} represents the distribution of the coordinates of the bodies in the simulations.
	
\begin{figure}[ht]
		\includegraphics[width=\linewidth]{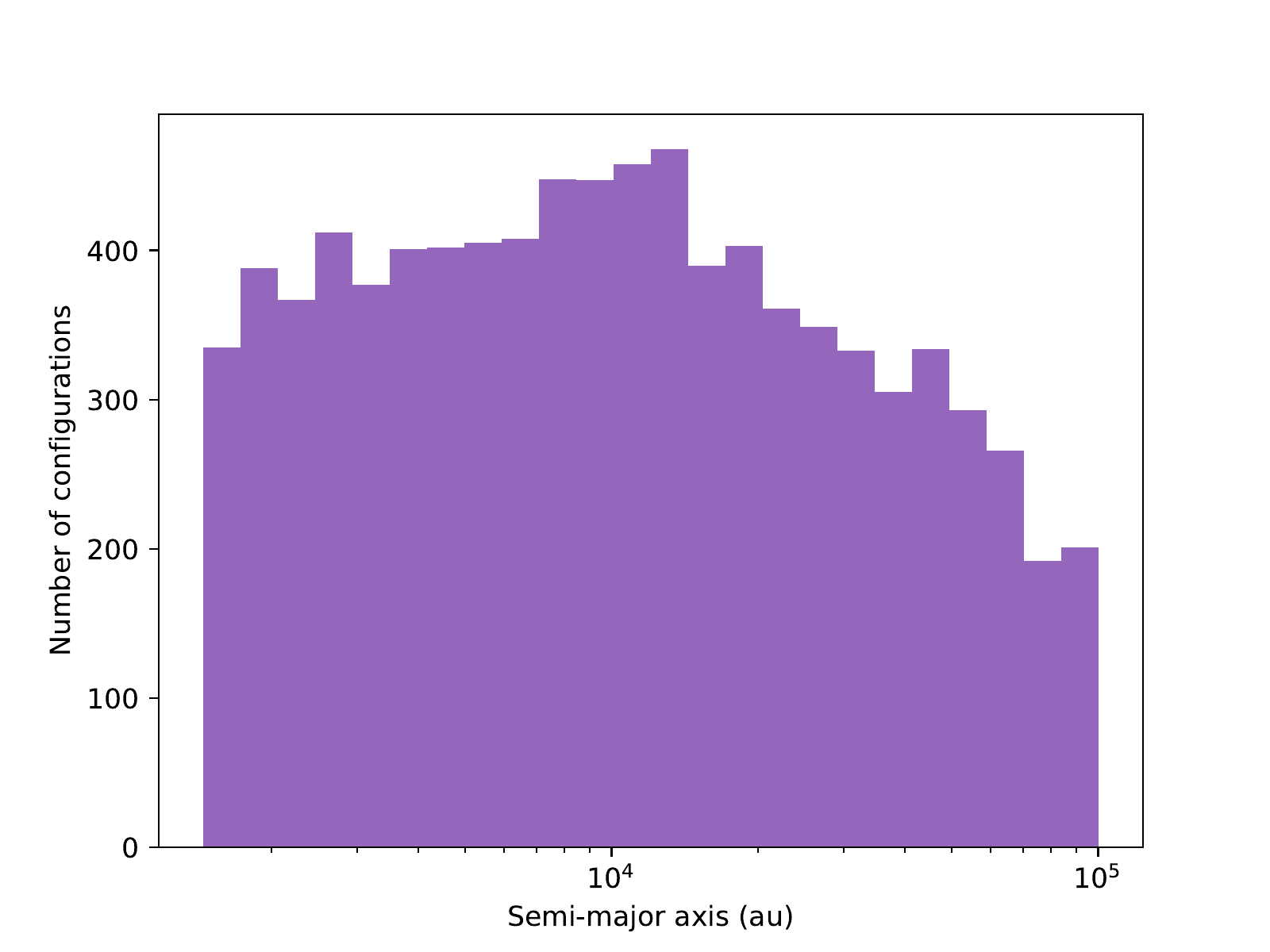}
		\includegraphics[width=\linewidth]{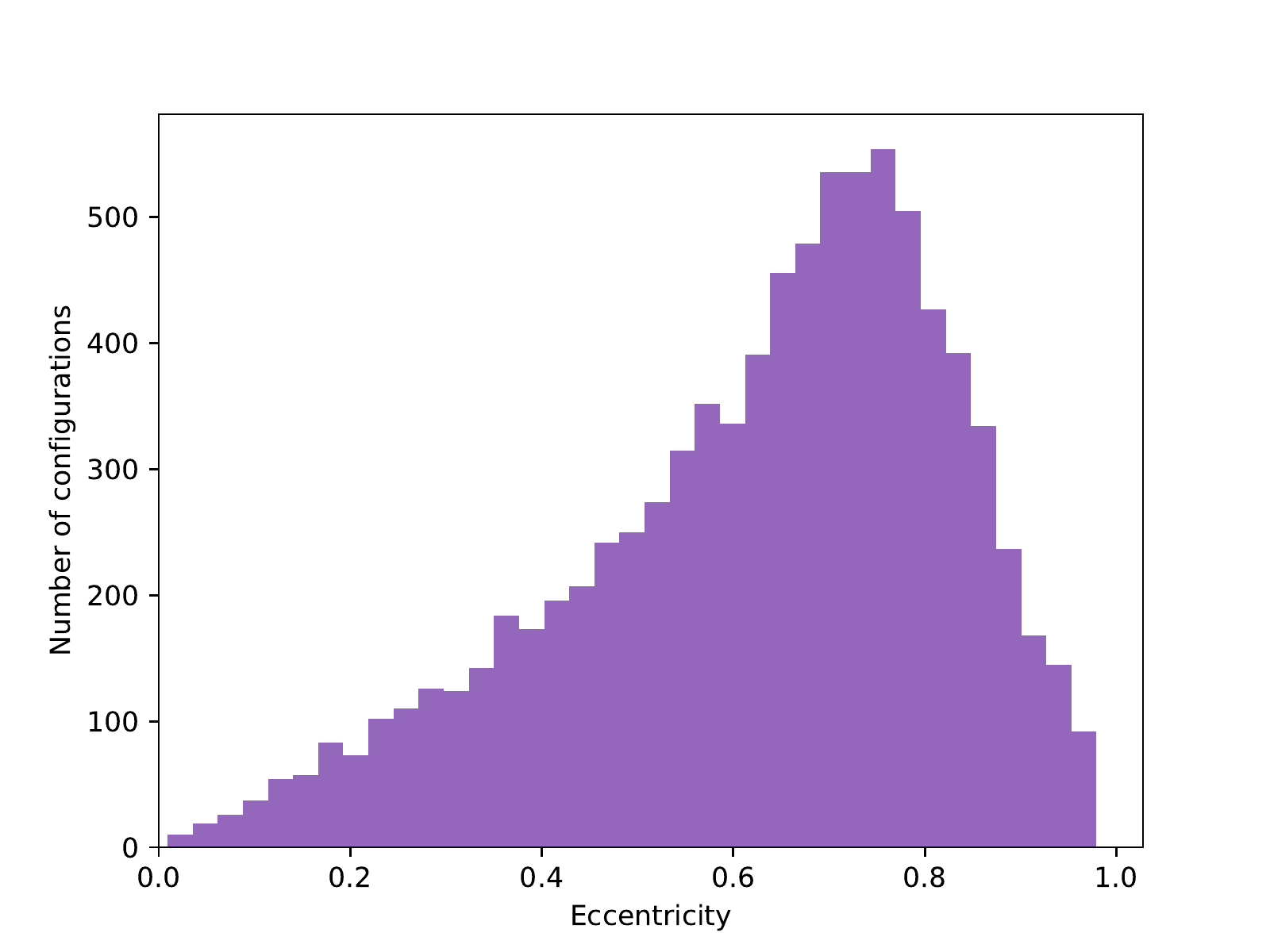}
		\caption{Semi-major axis and eccentricity distributions for the relative orbit of the two perturbers HIP 59716 and HIP 59721, assuming they are bound.}\label{fig:config}
	\end{figure}

\begin{figure}[h]
		\includegraphics[width=\linewidth]{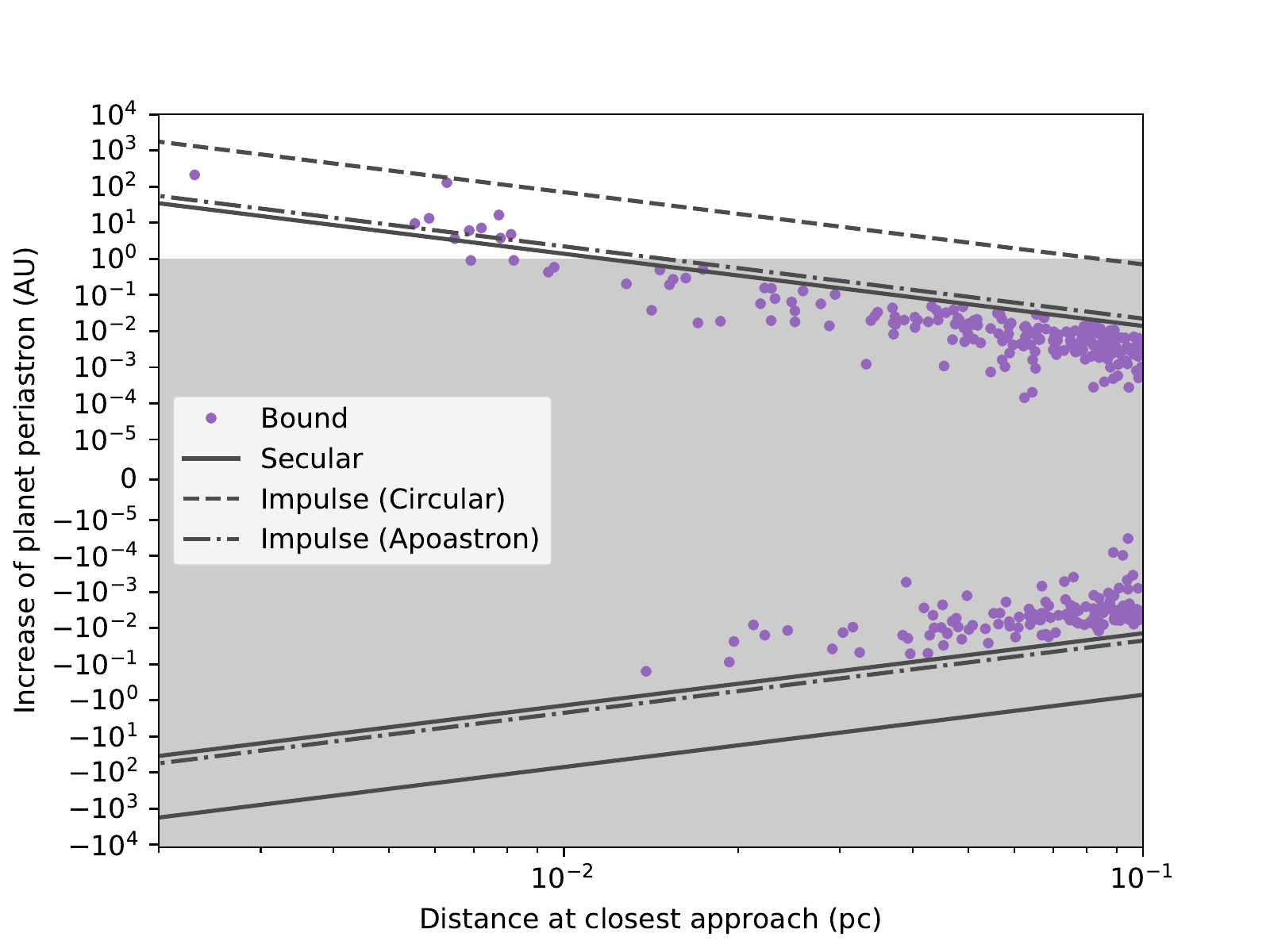}
		\caption{Periastron increase with respect to the distance at closest approach, from N-body simulations (dots) and theoretical approaches (lines), for the closer fly-bys, for an initial planetary apoastron of 1,000 au, in the case where the two perturbers are bound.}\label{fig:bound}
	\end{figure}

	Fig. \ref{fig:config} and \ref{fig:bound} describe the case where the two perturbers are bound. The coordinates of the bodies are drawn from the observational constraints with the same process that for the non-bound case, but we discarded the configurations where the eccentricity of the relative orbit is greater than 1. The resulting semi-major axis and eccentricity distributions are presented here, along with the effect of the fly-bys on the planet periastron, which is very similar to the non-bound case.
	
\end{document}